\documentclass[prb,aps,amssymb,twocolumn]{revtex4-1}
\usepackage{amsmath}
\usepackage{amssymb}
\usepackage{amsthm}
\usepackage{amsfonts}
\usepackage{listings}
\usepackage{enumerate}
\usepackage{latexsym}
\usepackage{color}
\usepackage[colorlinks=true,urlcolor=blue,citecolor=blue,linkcolor=blue]{hyperref}
\usepackage{psfrag}

\usepackage{bm}
\usepackage{graphicx}
\usepackage{subfigure}

\newcommand{\beq}{\begin{equation}}
\newcommand{\eneq}{\end{equation}}

\newcommand{\bal}{\begin{align}}
\newcommand{\eal}{\end{align}}

\input{epsf}

\begin{document}

\tolerance 10000

\newcommand{\vk}{{\bf k}}

\title{Topological nodal-line semimetals in ferromagnetic rare-earth-metal monohalides}

\author{Simin Nie$^{1}$}
\email{smnie@stanford.edu}
\author{Hongming Weng$^{2,3}$}
\author{Fritz B. Prinz$^{1,4}$}

\affiliation{${^1}$Department of Materials Science and Engineering, Stanford University, Stanford, California 94305, USA}
\affiliation{${^2}$Beijing National Laboratory for Condensed Matter Physics, and Institute of Physics,
Chinese Academy of Sciences, Beijing 100190, China}
\affiliation{${^3}$Collaborative Innovation Center of Quantum Matter, Beijing 100190, China}
\affiliation{${^4}$Department of Mechanical Engineering, Stanford University, Stanford, California 94305, USA}

\date{\today}

\begin{abstract}
Topological semimetals, extending the topological classification from insulators to metals, have greatly enriched our understanding of topological
states in condensed matter. This is particularly true for topological nodal-line semimetals (TNLSs). In the present paper, we identify layered materials
as promising candidates for hosting TNLSs. Based on first-principles calculations and effective model analysis, we propose that layered ferromagnetic
rare-earth-metal monohalides LnX (Ln=La, Gd; X=Cl, Br) exhibit long pursued topological phases. Specifically, single-layer LaX and single-layer GdX
are ideal two-dimensional (2D) Weyl semimetals and large-gap 2D quantum anomalous Hall insulators (QAHIs), with band gaps up to 61 meV,
respectively. In addition, 3D LaX and 3D GdX are TNLSs with a pair of mirror-symmetry protected nodal lines and 3D QAHIs, respectively. The
nodal lines in 3D LaX extending through the whole Brillouin zone (BZ) are fairly robust against strong spin-orbit coupling (SOC)  and located close to
the Fermi level, providing a novel platform toward exploring the exotic properties in nodal-line fermions as well as related device designs.
\end{abstract}
\maketitle

\section{introduction}
Nodal-line fermions\cite{burkov2011topological,fang2015topological,weng2016topological}, a new type of fermions beyond
Dirac\cite{young2012dirac,wang2012dirac,wang2013three} and Weyl fermions\cite{nielsen1983adler,wan2011topological,balents2011weyl},
have attracted ever-increasing interest in the field of condensed matter physics due to the absence of an elementary particle counterpart in
high-energy physics. Topological materials hosting nodal-line fermions (known as TNLSs) can be driven into various topological states, such
as Weyl semimetals\cite{yuHfC,du2017emergence}, Dirac semimetals\cite{kim2015dirac,yu2015topological}, topological
insulators\cite{srIo3,Nie03102017} and QAHIs\cite{LaXqahe,wu2017high}, by including SOC effect or breaking certain symmetry
(for example mirror symmetry). In TNLSs, the highest valence and the lowest conduction bands touch each other along 1D
symmetry-protected lines in the momentum space of the bulk, which can take many different forms, such as closed lines inside the
BZ\cite{wang2016body}, extended lines traversing the BZ\cite{chen2015nanostructured}, or nodal chain consisting
of connected lines\cite{bzduvsek2016nodal} \emph{etc.} The nodal lines carrying quantized $\pi$-Berry phase are protected by
crystal symmetry or the coexistence of time-reversal and inversion symmetries, and always lead to nearly flat drumhead-like
states with infinite density of states (DOS) on the boundary. These unique characteristics generate some exotic properties in
TNLSs: high-temperature surface superconductivity \cite{hsp1,volovik2015standard,heikkila2016flat}, unique Landau energy
level\cite{rhim2015landau}, and special long-range Coulomb interactions\cite{huh2016long}, \emph{etc.}

Although there has been great progress in theoretical prediction of TNLSs in real materials with small SOC interaction
\cite{wengC,xie2015new,chan2016,hirayama2017topological,jzhoublack,XuCaP3,chang2017realization,feng2018topological,wang2018topological},
the nontrivial nodal lines have been only confirmed in pure alkali earth metals by photoemission spectroscopy experiments\cite{li2016dirac}.
Owing to the fact that the non-negligible SOC can always gap out the nodal lines protected by inversion and time reversal symmetries, almost
all candidates evolve into either topological insulators or Dirac semimetals. Another route to TNLSs is to introduce the crystal symmetry (for example,
mirror symmetry or glide-mirror symmetry). However, very few materials are predicted to host nodal lines protected by the crystal symmetry
\cite{chen2015topological,bian2016topological}, which are robust against SOC. What's worse, most nodal lines are far away from the Fermi level and
not formed by band touching of the highest valence band and the lowest conduction band\cite{schoop2016dirac,fu2017observation}.
To the best of our knowledge, the signature of the existence of drumhead surface states is only observed in nonmagnetic $\text{PbTaSe}_2$
\cite{bian2016topological}, in which the existence of trivial Fermi pockets and some accidental nodal lines results in very complicated spectroscopic
and transport properties. In fact, more research is needed regarding the illustration of the direct relationship between the symmetry-protected nodal
lines and the drumhead surface states because of such complicated band structure.

The revealed TNLSs so far are extremely limited to nonmagnetic materials despite the discovery of various topological nontrivial states in magnetic
materials\cite{yu2010quantized,haijunCdOEuO,xu2011chern,chang2013experimental}. It is well known that magnetic order can significantly modify
the electronic structure and may give rise to the nontrivial nodal lines. Considering that the nodal lines in nonmagnetic materials can always be
gapped out under sufficiently large SOC even in $\text{PbTaSe}_2$, the realization of TNLSs in magnetic systems would be much more desirable
and significant. Here, we find that layered materials are very likely to host spinful nodal lines. Guided by this insight, we show that some long pursued
topological phases can be achieved in layered ferromagnetic LnX. The interlayer binding energies of 3D LnX are weaker than that of graphite, which
means it may be able to make single-layer LnX in a very simple and efficient way. Our calculations show that single-layer LaX are ideal 2D Weyl
semimetals with Weyl nodes located close to the Fermi level, and single-layer GdX are promising large-gap 2D QAHIs with band gaps up to 61 meV.
More interestingly, 3D LaX are TNLSs with a pair of spinful nodal lines nearly at the Fermi level, and 3D GdX are rare 3D QAHIs\cite{halperin1987possible}. In contrast
to conventional nodal lines, the spinful nodal lines with quantized $\pi$-Berry phase are protected by mirror symmetry with respect to the $xoz$ plane,
and can only be removed when they meet in the momentum space\cite{hyart2017two}. The topological phases in LnX, especially the TNLSs in 3D
LaX, pave a new way for studying the corresponding exotic properties in condensed matter systems.

\section{Calculation methods}

The full-potential linearized augmented plane-wave method implemented
in WIEN2K package\cite{blaha2002wien2k} is employed to perform the first-principles
calculations, which are cross-checked with the projector augmented wave method
implemented in Vienna ab initio simulation package\cite{kresse1996efficiency,kresse1996efficient}.
The exchange and correlation potential is treated within the local-density approximation (LDA) of Perdew- and
Wang-type\cite{LDA}. SOC is taken into account as a second variational step
self-consistently. The $k$-point grids of the BZ with 18 $\times$ 18 $\times$ 18
and 20 $\times$ 20 $\times$ 1 are used in the self-consistent calculations of 3D LnX
and single-layer LnX, respectively. The radii of the muffin-tin sphere $R_{MT}$ in
the calculations are 2.5 bohrs for La, Gd, and Br and 2.39 bohrs for Cl, respectively.
The LDA+U method\cite{liechtenstein1995density} is carried out to properly treat
the correlation effect in LnX with U = 5 eV and U = 6 eV for $d$ orbitals of La
and $f$ orbitals of Gd, respectively. The $s$ and $d$ orbitals of Ln are used to
construct the maximally localized Wannier functions (MLWFs) \cite{marzari2012maximally}, which are then used to calculate the
boundary states by an iterative method\cite{sancho1984quick,sancho1985highly,wu2017wanniertools}.

The strength of van der Waals forces holding the layers of bulk can be described by the interlayer
binding energy, which is defined as
\begin{eqnarray}
	&&E_b=-\frac{E_{\text{bulk}}-N \cdot E_{\text{layer}}}{N\times S_{\text{layer}}}\nonumber
\label{Binding}
\end{eqnarray}
where $E_{\text{bulk}}$ and $ E_{\text{layer}}$ are the total energies of a bulk unit cell and a
single-layer unit cell, respectively. $N$ is the number of layers in the bulk unit cell. $S_{\text{layer}}$
is the area of the single-layer unit cell.

The Berry phase of a ring ($\ell$) piercing the nodal line (shown as black lines in Fig. \ref{fg2}(e)) is quantized
to be $\pi$ and can be defined as
\begin{eqnarray}
	&&\phi_\ell=\oint_\ell \mathbf{A(k)}\cdot d\mathbf{k}  \nonumber
\label{Binding}
\end{eqnarray}
where $\mathbf{A(k)}=-i\displaystyle{\sum_{n\in occ.}}\langle u_n(\mathbf{k})|\partial_{\mathbf{k}}|u_n(\mathbf{k})\rangle$
is the Abelian Berry connection and $|u_n(\mathbf{k}) \rangle$ are the Bloch states. The Berry phase is
defined in terms of modulo $2\pi$.

\begin{figure*}[tp]
\center
\includegraphics[clip,scale=0.11, angle=0]{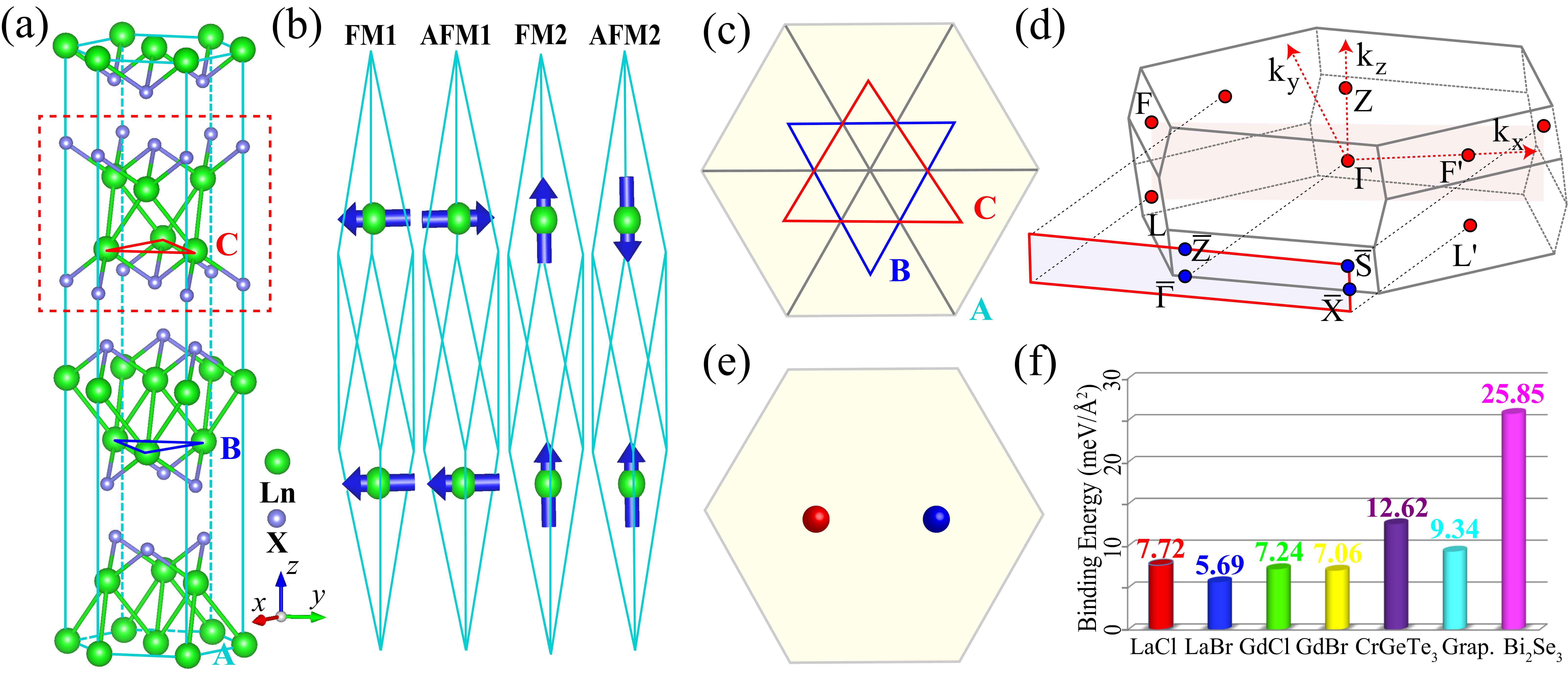}
\caption{(Color online) Crystal structure and topological phase evolution. (a) Crystal structure of 3D LnX. The green and light blue
balls represent Ln and X, respectively. The quadruple layer is indicated by the red dashed line. (b) Schematics of four different
collinear magnetic structures for LnX in the primitive cell. For simplicity, only Ln atoms are shown, while the nonmagnetic X atoms
are omitted. (c) Top view of LnX. Each unit cell consists of three quadruple layers stacked in an ABC-type trilayer pattern along
the $z$-direction. (d) The bulk BZ and the projected surface BZ for (010) surface of LnX. The grey plane indicates the $xoz$
plane. (e) Schematic of a 2D BZ and the boundary paths used to calculate the Berry phases. (f, g) Schematics of 2D Weyl
semimetal and 3D TNLS, respectively. (h) The interlayer binding energies of layered materials (LnX, CrGeTe$_3$,
Graphite and $\text{Bi}_2\text{Se}_3$).
}
\label{fg1}
\end{figure*}

\section{RESULTS AND DISCUSSION}
\subsection{Evolution of band crossings from 2D to 3D systems}

Both the theoretical proposal\cite{kane2005quantum,bernevig2006quantum} and the experimental realization\cite{konig2007quantum}
of topological nontrivial properties are based on 2D materials (2D topological insualtors) at the beginning, and 2D topological insulators
have been generalized to 3D topological insulators very soon\cite{fu2007topological}, which are classified into 3D strong topological
insulators and 3D weak topological insulators according to the interlayer coupling strength. Similar to the situation in the topological
insulators protected by time reversal symmetry, the nodal lines can also be found in the 3D materials constructed by stacking the
2D materials with twofold-degenerate band crossings (called Weyl nodes).
Although H. Nielsen and M. Ninomiya only shown that the no-go theorem holds on 1D and 3D lattices \cite{nielsen1981absence1,nielsen1981absence2}, the theorem is also applicable to 2D lattice with specified symmetry \cite{fang2015new,shiozaki2015z,fang2017rotation}, such as time reversal symmetry. Contrary
to the fairly robust Weyl nodes in 3D materials, the Weyl nodes in 2D materials are unstable. Thus, additional symmetry is needed to guarantee its existence, such as mirror symmetry.
Therefore, for simplicity, we only consider a pair of Weyl nodes in a 2D material with mirror symmetry $\hat{M}_y$, as shown in Fig. \ref{fg1}(e). Then, the
2D Weyl semimetal is stacked into a 3D layered material, which can be grouped into three different nontrivial semimetal classes depending
on the interlayer coupling strength and the symmetry.
For the symmetry-protected Weyl nodes stacked along a line preserving the symmetry, the Weyl nodes can evolve into two different classes (see more details in the Appendix A): Class1,
two nodal loops extending through the BZ for the layered materials with weak interlayer coupling; Class2,
closed loops or accidental nodal chain for the layered materials with strong interlayer coupling. On the other hand,
if the symmetry is broken on the stacking line, the layered material may become a 3D Weyl semimetal, which belongs to Class3.
Therefore, in consideration of the robust existence of nodal lines in layered materials,  it is much promising to search for the TNLSs in
this kind of material with proper symmetry.

\subsection{Crystal structure, magnetic configurations, and Nodal lines}

Guided by this idea, the nontrivial nodal lines are found in a class of topological materials LnX. LnX are well-known layered
materials and have been synthesized since the early 1980s\cite{araujo1981lanthanum,mattausch1980monohalogenide}.
Experimentally, they all crystalline in the hexagonal layered structure with space group  $R \bar{3}m$ (No. 166), as shown
in Fig. \ref{fg1}(a). The basic building block of LnX is a tightly bound quadruple layer, which consists of two
hexagonal rare-earth-metal (Ln) layers sandwiched between two hexagonal halogen (X) layers in the sequence of
X-Ln-Ln-X. The adjacent quadruple layers are stacked in an ABC-type trilayer pattern along the $z$-direction with
much weak van der Waals (vdW) interaction, as shown in Fig. \ref{fg1}(c).

Given that the interlayer distance is very close to 10 \AA, the interlayer binding strength of LnX should be much weaker than
that of most layered compounds. We have calculated the interlayer binding energies of LnX as well as three well-established
layered materials (graphite, $\text{Bi}_2\text{Se}_3$, and $\text{Cr}\text{Ge}\text{Te}_3$). The results are shown in
Fig. \ref{fg1}(f), which are consistent with previous results\cite{weng2014transition} and cross-checked with vdW density
functional (vdW-DF) calculations\cite{dion2004m}. It is easy to find that $\text{Bi}_2\text{Se}_3$ has the highest binding
energy among these seven materials and the interlayer bonding of $\text{Cr}\text{Ge}\text{Te}_3$ is as weak as that of
graphite, which is a main reason for the recent realization of 2D ferromagnetic $\text{Cr}\text{Ge}\text{Te}_3$
\cite{gong2017discovery}. More interestingly, the interlayer coupling of $\text{LnX}$ is even weaker than that of
graphite, suggesting single-layer $\text{LnX}$ can be easily produced by exfoliation methods.

\begin{figure*}[htp]
\center
\includegraphics[clip,scale=0.25, angle=0]{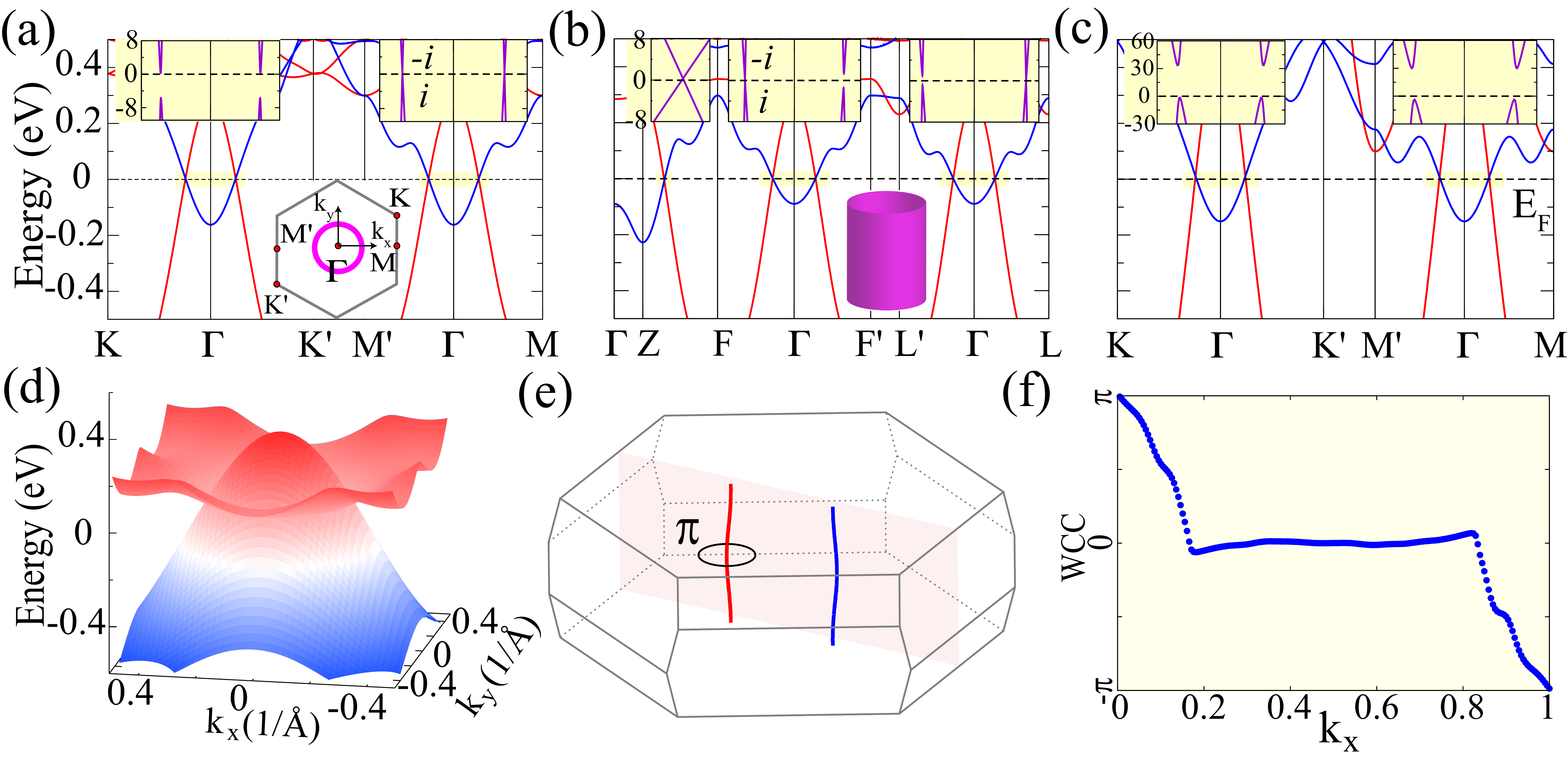}
\caption{(Color online) Electronic structures of LnX. (a, b, c) The LDA+U band structures of single-layer LaCl, 3D LaCl and single-layer
GdCl, respectively. The red and blue lines represent spin-up and spin-down states, respectively. The upper insets show the zoom-in
LDA+U+SOC band structures around the band crossing points. The lower insets of (a) and (b) show the schematics of the nodal line
and cylinder, respectively. The eigenvalues of mirror symmetry $\hat{M}_{y}$ are labeled. The energy unit of upper insets is meV
instead of eV.  (d) 3D band structure of 3D LaCl in the $xoy$ plane without the consideration of SOC. (e) Two nodal lines of 3D LaCl
are located in the $xoz$ plane, which is indicated as grey plane. (f) The evolution of Wannier charge centers for single-layer GdCl.}
\label{fg2}
\end{figure*}

Next, we consider the magnetic structure in LnX. Although magnetic structures and the transition temperatures of LnX have
not been reported yet, it is reasonable to expect magnetic order in LnX due to the monovalent Ln. Due to the fact that the
results of single-layer LnX are similar to that of 3D LnX, we only discuss the results of 3D LnX in the following (see more
details in the Appendix B). Here, by using LDA+Hubbard U+SOC (LDA+U+SOC), we calculate the total energies of four
different magnetic configurations for 3D LnX, including two ferromagnetic (FM1, FM2) and two antiferromagnetic (AFM1, AFM2)
configurations shown in Fig. \ref{fg1}(b), and the converged total energies are summarized in Table \ref{totenergy}.
The results clearly show that both 3D LaX and 3D GdX prefer a ferromagnetic ground state, lowering the total energy in
the range of dozens meV than the AFM configurations. More interestingly, 3D LaX and 3D GdX prefer FM1 with the spin
aligned along the $y$-direction and FM2 with the spin aligned along the $z$-direction, respectively, in spite of the small
energy difference between FM1 and FM2 (2.09/1.52 meV and 0.72/1.08 meV for 3D LaCl/3D LaBr and 3D GdCl/3D GdBr, respectively).
In addition, we have checked the magneto-crystalline anisotropy energy within the $xoy$ plane for LaX. The magneto-crystalline anisotropy energy is of the order of 0.0001 meV, indicating LaX are soft magnetic materials with an easy magnetization axis in the $xoy$ plane. Experimentally, an external magnetic field can be used to realize the FM1 state and the topological semimetals in LaX. So we only focus the topological properties of LaX with FM1 state in the following discussion.

The point group for the nonmagnetic materials with space group $R \bar{3}m$ is the $D_{3d}$ symmetry group, which has
three generators: threefold rotational symmetry about the $z$-axis ($\hat{C}_3^z$), inversion symmetry ($\hat{I}$), and mirror
symmetry with respect to the $xoz$ plane ($\hat{M}_{y}$). For 3D LaX, the ferromagnetic order (FM1) breaks  $\hat{C}_3^z$
and the little group is reduced from $D_{3d}$ to $C_{2h}$ with generators $\hat{I}$ and $\hat{M}_{y}$ only.
However, for 3D GdX, FM2 breaks $\hat{M}_{y}$ and the little group becomes $C_{3i}$ with generators  $\hat{I}$ and $\hat{C}_3^z$ only.
The FM ground states with different symmetries lead to different band structures and topological phases in LnX, as will be shown below.

Since LaBr has almost the same results as LaCl, we choose LaCl as an example in the following demonstration. The calculated
band structures of single-layer LaCl by LDA+U are shown in Fig. \ref{fg2}(a), which show a very deep band inversion at
$\Gamma$ point.
When SOC is ignored, the band inversion results in a nodal line circled around $\Gamma$ point, as
shown in the lower inset of Fig. \ref{fg2}(a). The nodal line is mainly
contributed by Ln $|d_{z^2}\rangle$, $|d_{x^2-y^2}\rangle$ and $|d_{xy}\rangle$ orbitals (see
more details in the Appendix C).
Then, when SOC is taken into consideration, the nodal line is gapped out, except
the band crossings of $k$ points on the $M-\Gamma-M'$ line, decaying into two Weyl nodes, as shown in the upper insets of
Fig. \ref{fg2}(a). As discussed above, the ideal Weyl nodes in 2D LaCl with the nodes located close to the Fermi level may lead to
nodal lines in 3D LaCl due to the extremely weak interlayer coupling. In order to confirm the speculation, we calculate the
band structures of 3D LaCl, as shown in Fig. \ref{fg2}(b) and \ref{fg2}(d). Without the consideration of SOC, there is a cylinder
centered around $\Gamma$ point that originates from the nodal line of single-layer LaCl, as shown in the lower inset of
Fig. \ref{fg2}(b). Considering SOC, the cylinder is gapped out everywhere except the band crossings of $k$ points in the
$xoz$ plane (i.e., 3D LaCl belongs to the Class1), as shown in the upper insets of Fig. \ref{fg2}(b). The calculations of the
eigenvalues of mirror symmetry $\hat{M}_y$ indicate that the two crossing bands have opposite mirror eigenvalues $\pm i$,
which means the nodal lines are protected by mirror symmetry $\hat{M}_y$. It is worth noting that the extended
nodal lines in 3D LaCl always appear in pairs, as shown in Fig. \ref{fg2}(e), which is different from the conventional nodal lines.
They can only be annihilated without breaking $\hat{M}_y$ when they meet in the momentum space.

\subsection{k$\cdot$p analysis and drumhead-like surface states}

As discussed above, the little group at $\Gamma$ point for 3D LaX is $C_{2h}$ with generators $\hat{I}$ and $\hat{M}_{y}$ only.
In the following, we can prove by a simple $k\cdot p$ analysis in the vicinity of $\Gamma$ point that these two symmetries are
enough to guarantee the existence of a pair of nodal lines around $\Gamma$ point in 3D LaX. An
effective low-energy 2$\times$2 model capturing the nodal lines can be in general written in the form
\begin{eqnarray}
H^{\Gamma}(\mathbf{k})=\sum_{i=x,y,z} d_{i}(\mathbf{k})\sigma_{i}
\label{kp1}
\end{eqnarray}
where $\sigma_{i}$ are three Pauli matrices; $d_{i}(\mathbf{k})$ are real functions; $\mathbf{k}=(k_x, k_y, k_z)$ is the momentum vector
relative to $\Gamma$ point. The identity matrix $\sigma_0$ just shifts the Fermi level and is ignored in Eq. \ref{kp1}. The two crossing
bands around $\Gamma$ point in 3D LaX have the same parity, which indicates that the inversion symmetry $\hat{I}$ is represented
by $\hat{I}=\sigma_0$. Then, the inversion symmetry requires that
\begin{eqnarray}
\hat{I}H^{\Gamma}(\mathbf{k})\hat{I}^{-1}=H^{\Gamma}(-\mathbf{k})
\label{kp2}
\end{eqnarray}
which means $d_{i}(\mathbf{k})$ are even functions of $\mathbf{k}$ and can be up to the second order of $\mathbf{k}$ generally written as
\begin{eqnarray}
d_{i}(\boldsymbol{k})=c^i_{0}+c^i_{1}k_{x}^{2}+c^i_{2}k_{y}^{2}+c^i_{3}k_{z}^{2} \nonumber \\
                                 +c^i_{4}k_{x}k_{y}+c^i_{5}k_{y}k_{z}+c^i_{6}k_{x}k_{z}
\label{dk1}
\end{eqnarray}
where $c^i_n$, $n=0,...,6$, are parameters characterizing the band structure. In addition, the two crossing bands have opposite
mirror eigenvalues, which means the mirror symmetry $\hat{M}_{y}$ can be represented by $\hat{M}_{y}=\sigma_{z}$. Therefore,
another constraint on $H^{\Gamma}(\mathbf{k})$ placed by the mirror symmetry is
\begin{eqnarray}
\sigma_{z}H^{\Gamma}(k_{x},k_{y},k_{z})\sigma_{z}=H^{\Gamma}(k_{x},-k_{y},k_{z})
\label{kp3}
\end{eqnarray}
which leads to
\begin{eqnarray}
d_{x/y}(k_{x},k_{y},k_{z}) &=&-d_{x/y}(k_{x},-k_{y},k_{z}) \label{dk2} \\
d_{z}(k_{x},k_{y},k_{z})   &=&d_{z}(k_{x},-k_{y},k_{z}) \label{dk3}
\end{eqnarray}
Considering the constraints on $d_i(\mathbf{k})$ (Eq. \ref{dk1}, \ref{dk2}, and \ref{dk3}), we can get
\begin{eqnarray}
d_{x,y}(k_{x},k_{y},k_{z})\!&=&\!c^{x,y}_{4}k_{x}k_{y}+c^{x,y}_{5}k_{y}k_{z} \label{dk4} \\
d_{z}(k_{x},k_{y},k_{z})\!&=&\!c^z_{0}\!+\!c^z_{1}k_{x}^{2}\!+\!c^z_{2}k_{y}^{2}\!+\!c^z_{3}k_{z}^{2}\!+\!c^z_{6}k_{x}k_{z} \label{dk5}
\end{eqnarray}

Then, we discuss the constraints on $c^i_n$ and the emergence of nodal lines due to the band inversion at $\Gamma$ point in
3D $\text{LaX}$. It is easy to find that the band inversion requires $c^z_{0}<0$. In addition, the band order along the $k_{x}$
direction shows a transition from inverted to normal order, which indicates $c^z_{1}>0$. However, the two bands are always
inverted along the $k_{z}$ direction, which indicates $c^z_{3}<0$. The signs of $c^z_n$  are important to give rise to nodal
lines, which can be easily proved by following simple discussion. The appearance of the band crossing points in the $k_{y}=0$ plane requires
\begin{eqnarray}
c^z_{0}+c^z_{1}k_{x}^{2}+c^z_{3}k_{z}^{2}+c^z_{6}k_{x}k_{z}=0
\label{dk6}
\end{eqnarray}
Given $4*c^z_{1}\cdot c^z_{3}<0<{(c^z_{6})}^{2}$, the solutions, the band crossing points, form a hyperbola in the $k_{y}=0$ plane.

\begin{figure}[t]
\center
\includegraphics[clip,scale=0.36, angle=0]{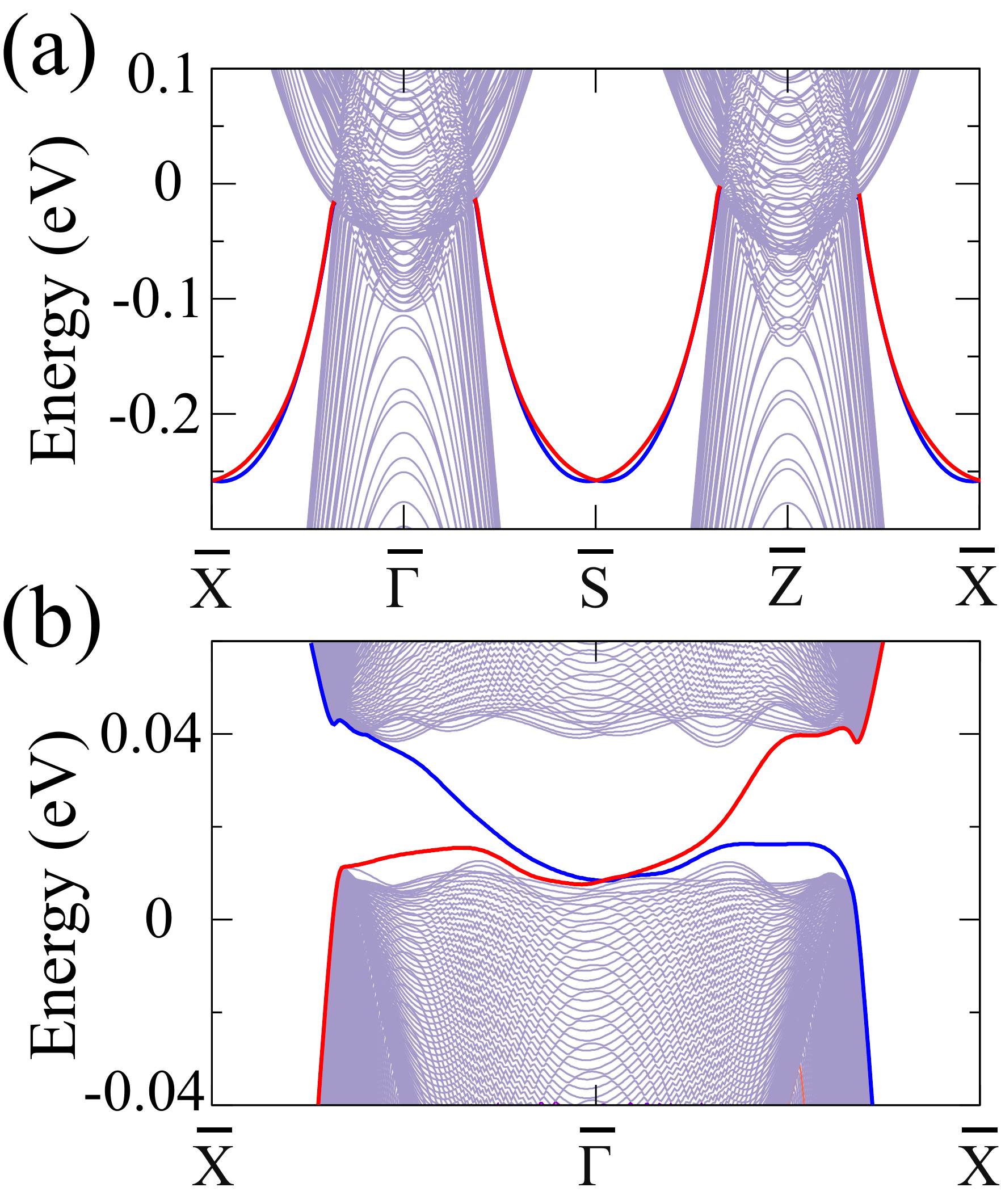}
\caption{(Color online) Boundary states. (a)  The surface states of 3D LaCl in the (010) plane.
(b) The edge states of single-layer GdCl. The red and blue bands represent the boundary states located at
opposite boundaries.}
\label{fg3}
\end{figure}

Since the existence of nontrivial boundary states is a hallmark of TNLSs, we have calculated the surface states of 3D LaCl by
constructing the Green's functions based on the MLWF method, as shown in Fig. \ref{fg3}(a). It is obvious that the nontrivial
drumhead-like surface states can be found in the bulk band gap, which confirms that 3D LaCl is a TNLS clearly. The drumhead-like
 surface states should be easily observed in the future experiments, such as angle-resolved photoelectron spectroscopy (ARPES)
 or scanning tunnel microscope (STM).

\subsection{Large-gap 2D QAHIs in GdX}

Although GdX share the same hexagonal structure as LaX, the ground magnetic state of GdX is FM2, as discussed above,
which results in different electronic structures in GdX. GdX have almost the same results, and thus, we choose GdCl as an
example in the following illustration. The band structure of single-layer GdCl without SOC is shown in Fig. \ref{fg2}(c), which
shows a similar dispersion to single-layer LaCl at a quick glance. However, unlike the single-layer LaCl,
the nodal line circled around $\Gamma$ point is fully gapped out in the presence of SOC, as shown in the upper insets of Fig. \ref{fg2}(c).
In spite of the unbroken threefold rotation symmetry $\hat{C}_3^z$, the magnetic order (FM2) breaks the mirror symmetry $\hat{M}_{y}$,
which means that the nodal line loses the protection and is gapped out with a large band gap (39.4 meV and 61 meV for single-layer
GdCl and single-layer GdBr, respectively).

The band gap opening at the nodal line always induce topological nontrivial phase, which can be identified by the
calculation of the evolution of Wannier charge centers (WCC), as shown in Fig. \ref{fg2}(f). It is clearly shown that single-layer
GdCl is a QAHI with Chern number $C=-1$. Furthermore, based on the MLWF method,
we carry out the calculation of the edge states, as shown in Fig. \ref{fg3}(b), in which each edge contains a topologically
protected chiral edge state in the bulk band gap. Therefore, single-layer GdCl is a large-gap 2D QAHI. When a 2D QAHI is
stacked into a 3D material, the 3D material can evolve into two different nontrivial phases according to the interlayer coupling
strength: 3D QAHI and Chern semimetal  also called Weyl semimetal (see more details in the Appendix D). Our calculations
show that 3D GdCl is a 3D  QAHI due to the extremely weak interlayer binding energy. Compared with the
well-known Chern semimetal, 3D QAHI is studied very few due to very rare candidates. So 3D GdCl provides an ideal platform for future study.

\begin{figure*}[tp]
\centering
\includegraphics[width=6.0in]{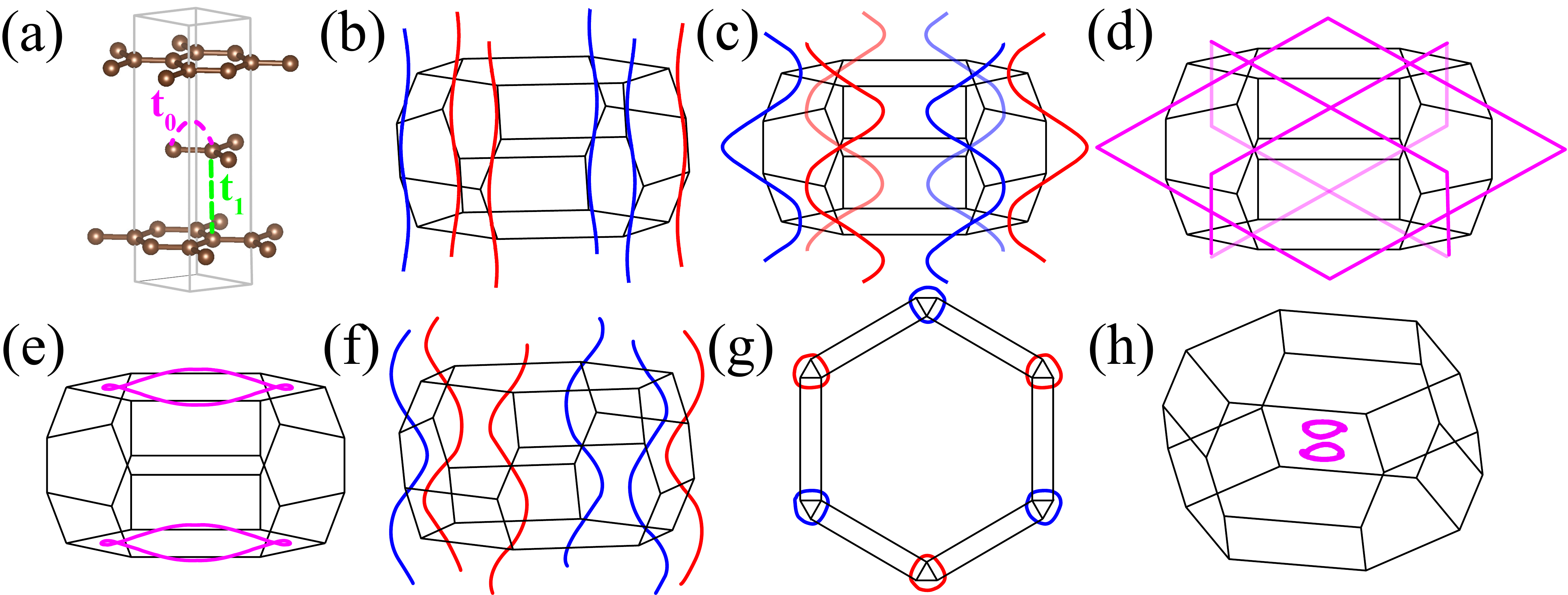}
      \caption{(Color online) (a) The crystal structure of the ABC-stacked graphene. $t_0$ and $t_1$ are the nearest intralayer and interlayer hoppings, respectively.  (b-e) The evolution of the nodal lines with different $t_1/t_0$: (b) $t_1/t_0=0.1$, (c) $t_1/t_0=0.9$, (d) $t_1/t_0=1$, (e) $t_1/t_0=1.5$. (f-h) The nodal lines in the ABC-stacked graphene calculated by first-principles method. (f, g) The side view and top view of the nodal lines in the ABC-stacked graphene with interlayer distance 2.53 \AA, respectively. (h) The nodal lines in the ABC-stacked graphene with interlayer distance 1.63 \AA. }
\label{graphene}
\end{figure*}

\section{Conclusion}

To summarize, we show that it is much desirable to realize spinful nodal lines in 3D layered materials constructed by stacking 2D Weyl
semimetals. Following this methodology as a guideline, we propose that fruitful topological phases can be found in layered ferromagnetic
LnX by using the first-principles calculations and effective model analysis. More specifically, 3D LaX exhibit the long perceived spinful
nodal lines extending through the whole BZ, which are protected by mirror symmetry $\hat{M}_y$. The spinful nodal lines appearing in
pairs are fairly robust against SOC and can only be removed by moving them together in the momentum space, which is totally different
from the existing nodal lines. In addition, 3D GdX, single-layer LaX and single-layer GdX are rare 3D QAHIs,
ideal 2D Weyl semimetals and large-gap 2D QAHIs, respectively. The topological phases in LnX open a new path to realize exotic
fermions, especially the nodal-line fermions, and to explore many novel properties.

\begin{acknowledgments}

The authors thank Quansheng Wu, Zhijun Wang and Zhida Song for useful discussions. F. B. P. and S. N. were supported by
Stanford Energy 3.0. H. W. was supported by the National Key Research and Development Program of China under grant
No. 2016YFA0300600, and NSFC under grant number 11421092.

\end{acknowledgments}

\appendix
\section{Nodal lines in layered materials}

\begin{figure*}[tp]
\centering
\includegraphics[width=6.5in]{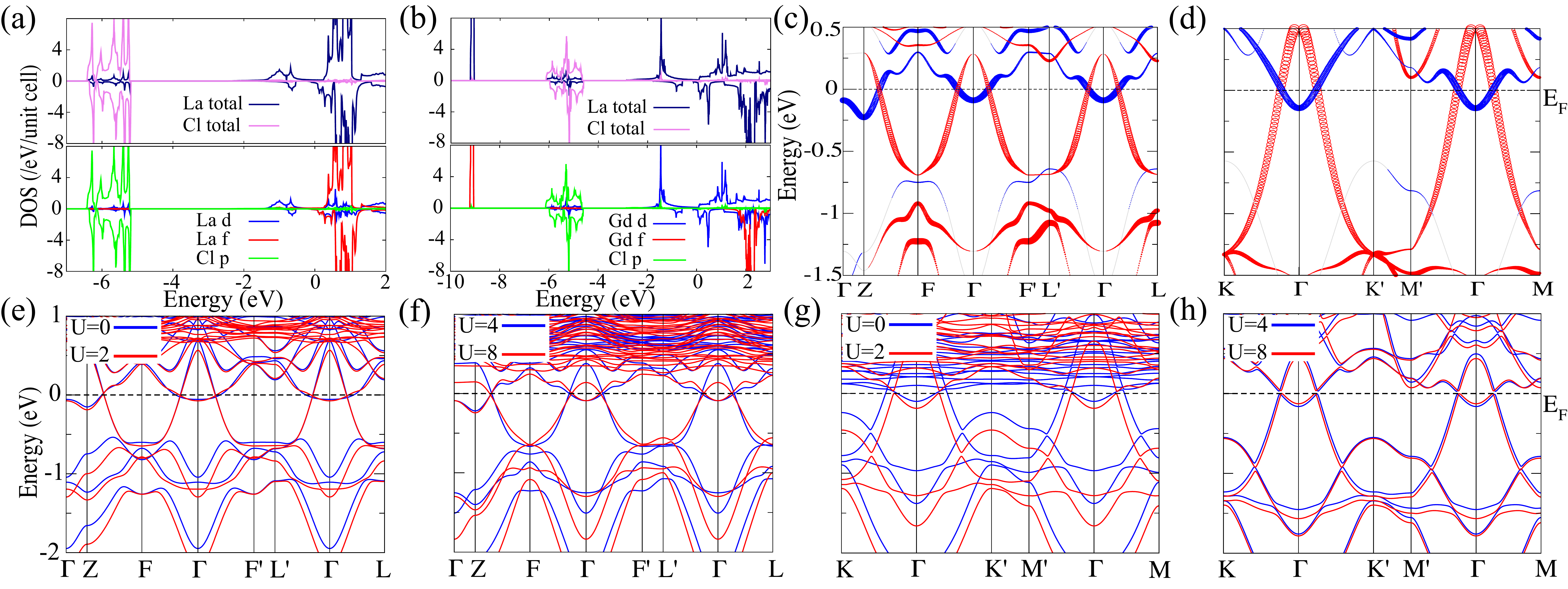}
      \caption{(Color online) The LDA+U-calculated projected density of states of 3D LaCl (a) and 2D GdCl (b).
      The fatted band structures of 3D LaCl (c) and 2D GdCl (d) by LDA+U calculations. The blue and red dots represent the weights of
      Ln $|d_{z^2}\rangle$ and $|d_{x^2-y^2}\rangle+|d_{xy}\rangle$, respectively. The LDA+U+SOC band
      structures of 3D LaCl (e, f) and 2D GdCl (g, h).
      }
\label{bands}
\end{figure*}

In order to clearly illustrate the evolution of band crossings from 2D to 3D systems, we choose an ABC-stacked graphene as an example
containing two inequivalent sublattices with only one $p_z$-type orbital for each sublattice, as shown in Fig. \ref{graphene}(a). To
simplify our illustration, only the nearest intralayer  ($t_0$)  and interlayer hoppings  ($t_1$)  have been taken into consideration,
and a 2$\times$2 Slater-Koster\cite{sktb} tight-binding (TB) model under the crystal symmetry restrictions can be written down as follows:
     \begin{align}
        H(\boldsymbol{k}) =
        &\left( \begin{array}{cccc}
           0&t_{0}\delta(\mathbf{k})+t_1 \lambda(\mathbf{k})\\
* & 0
        \end{array}\right)  \label{H0k} \\
        \nonumber
    \end{align}
where $\delta (\mathbf{k})=e^{2\pi i (\frac{1}{3}k_x+\frac{1}{3}k_y-\frac{2}{3}k_z)} +e^{2\pi i (\frac{1}{3}k_x-\frac{2}{3}k_y+\frac{1}{3}k_z)}+e^{2\pi i
(-\frac{2}{3}k_x+\frac{1}{3}k_y+\frac{1}{3}k_z)}$, $\lambda(\mathbf{k})=e^{2\pi i (\frac{1}{3}k_x+\frac{1}{3}k_y+\frac{1}{3}k_z)} $. $k_{x,y,z}$ are
defined with respect to the reciprocal lattice vectors.

It is well known that there are two band crossings at the Fermi level in the momentum space of graphene\cite{sarma2011electronic}, which
are protected by space inversion $\hat{I}$ and time reversal ($\mathcal{T}$) symmetries and constrained  at $\mathbf{K}$ and $\mathbf{K}'$
points respectively by threefold rotation symmetry about the $z$-axis ($\hat{C}_3^z$). When 2D graphene layers are stacked in an ABC-type
pattern and the interlayer coupling is very weak ($t_1/t_0=0.1$), the band crossing points located at $\mathbf{K}$ and $\mathbf{K}'$ points
evolve into two extended nodal lines, respectively, as shown in Fig. \ref{graphene}(b). As discussed in the main text, the nodal lines can
only be annihilated when they meet in the momentum space, which can be easily seen by increasing the interlayer coupling strength. When
$t_1/t_0=0.9$, the nodal lines are bended greatly and nearly touch each other, as shown in Fig. \ref{graphene}(c). At the critical point
( $t_1/t_0=1$), the nodal lines connect with each other and become a nodal chain\cite{bzduvsek2016nodal}, as shown in Fig. \ref{graphene}(d).
If we continue to increase $t_1/t_0$ to $1.5$, the nodal chain becomes a closed nodal line, as shown in Fig. \ref{graphene}(e). The closed
nodal line can be gapped out  when $t_1/t_0=3$. Therefore, the nodal lines in the layered materials are very robust, and it is much promising
to search for the TNLSs in this kind of materials.

In order to check whether this simple TB model can capture the main physics of the evolution, we have calculated the electronic structures of the
ABC-stacked graphene by the first-principles method, as shown in Fig. \ref{graphene}(f-h). When the interlayer distance equals to 3 \AA, we get
two nearly straight nodal lines, which are similar to the results in Fig. \ref{graphene}(b). Then we decrease the interlayer distance to increase
the interlayer binding strength. When the interlayer distance equals to 2.53 \AA, the straight nodal lines are bended spirally, as shown in
Fig. \ref{graphene}(f). If we view the nodal lines along $z$-direction (top view), the nodal lines form a ring circled around $\mathbf{K}$ and
$\mathbf{K}'$ points, respectively, as shown in Fig. \ref{graphene}(g). The size of the rings has a monotonically increasing relationship with
the interlayer coupling strength. At certain interlayer distance the two nodal lines touch each other, and the extended lines become closed
lines in the BZ after the critical point. As shown in Fig. \ref{graphene}(h), there are two closed nodal lines in the BZ when the interlayer
distance equals to 1.63 \AA. Although the number of the closed nodal lines are not equal to the results obtained from the TB model, the evolution
of the nodal lines calculated by the TB and first-principles methods are qualitatively similar.

 \begin{table*}[tp]
\caption{The converged total energies (unit: eV) of four different magnetic structures (FM1, FM2, AFM1, and AFM2) for LnX.}
  \begin{tabular}{ccc c c c c c c c}
\hline
        Config.    &  3D LaCl &  2D LaCl     & 3D LaBr      & 2D LaBr      & 3D GdCl      & 2D GdCl & 3D GdBr      &2D GdBr  \\
\hline
        FM1        & -16.63271& -16.51641 & -14.77393 & -14.70523  & -38.71503 & -38.76298 & -37.40773 & -37.42195
\\
        FM2        & -16.63062& -16.51447 & -14.77241 & -14.70362  & -38.71575 & -38.76376 & -37.40881 & -37.42294
\\
        AFM1       & -16.61171& -16.49404 & -14.74455 & -14.67597  & -38.57725 & -38.62517 & -37.21060 & -37.22607
\\

        AFM2       & -16.60992& -16.49223 & -14.74311 & -14.67425  & -38.57707 & -38.62562 & -37.21102 & -37.22683
\\
\hline
\end{tabular}
\label{totenergy}
\end{table*}

~\\

\section{Magnetic configurations and total energy calculations in LnX}

Considering the partial occupations of localized orbitals of Ln, we propose four different collinear magnetic
structures (FM1, FM2, AFM1, and AFM2) of LnX, as shown in Fig. \ref{fg1}b. As we can see, the magnetic moments of
collinear FM1/AFM1 and FM2/AFM2 are aligned along the $y$- and the $z$-directions, respectively. Then, the
first-principles total energy calculations are preformed based on the four magnetic structures, and the
results are shown in Table \ref{totenergy}, which clearly indicates that single-layer LnX have the same
ground states as that of 3D LnX. To be specific, the total energies of the AFM states are about 19, 27, 138,
and 198 meV/u.c. higher than that of the FM states for LaCl, LaBr, GdCl and GdBr, respectively. Moreover,
although the FM states have very close total energies, LaX and GdX prefer different FM states, \emph{i.e.} FM1,
and FM2, respectively. For 3D LaCl/3D LaBr (single-layer LaCl/single-layer LaBr), FM1 states further lower
the total energy about 2.09/1.52 meV/u.c. (1.94/1.61 meV/u.c.) compared with the FM2 states, while the total
energies of FM2 states are about 0.72/1.08 meV/u.c. (0.78/0.99 meV/u.c.) lower than that of FM1 states for
3D GdCl/3D GdBr (single-layer GdCl/single-layer GdBr). As discussed in the main text, the different magnetic
orders break different crystal symmetries and give rise to fruitful topological nontrivial phases in LnX.

~\\
\section{Electronic structures of LnX}

The projected density of states of 3D LaCl and 2D GdCl calculated by LDA+U are shown in Fig. \ref{bands}(a) and \ref{bands}(b), respectively.
As we can see, Cl $p$ orbitals of  3D LaCl and 2D GdCl are mainly contributed to the states in the energy intervals (-6.5 eV to -5.1 eV) and
(-6.2 eV to -4.6), respectively. La $f$ orbitals are unoccupied for both spin up and spin down channels, while the spin up components of
Gd $f$ orbitals are fully occupied and locate from -9.2 eV to -9.1 eV. For the bands around the Fermi level (0 eV), it is clearly shown
that they are dominated by the Ln $d$ and Ln $f$ orbitals. In order to get detail orbital character of the crossing bands, we performed
the fatted band structure calculations, as shown in Fig. \ref{bands}(c) and \ref{bands}(d). As pointed out in the main text, the point group of nonmagnetic
LnX is $D_{3d}$, five degenerate $d$ orbitals in the trigonal field effect are split into one singlet ($d_{z^2}$) and two doublets
(one is $d_{x^2-y^2}$ and $d_{xy}$, and the other one is $d_{xz}$ and $d_{yz}$). It is obvious that there exists an obvious weight
exchange near the Fermi level and the two crossing bands are dominated by  Ln $|d_{z^2}\rangle$ and
$|d_{x^2-y^2} \rangle+|d_{xy}\rangle$, whose weights are represented by the size of the blue and red dots, respectively.
By combining these atomic orbitals, the explicit form of the two low-energy states can be written down as $r_1(|Ln_1 d_{z^2}\rangle-|Ln_2 d_{z^2}\rangle)$ and
a combination of $r_2(|Ln_1 d_{x^2-y^2}\rangle-|Ln_2 d_{x^2-y^2}\rangle)$ and
$r_3 (|Ln_1 d_{xy}\rangle-|Ln_2 d_{xy}\rangle)$, where $r_1$, $r_2$ and $r_3$ are material-dependent coefficients.
For example, for 3D LaCl, $r_1$, $r_2$ and $r_3$ equal to 0.359, 0.342 and 0.342, respectively.

In addition, the nontrivial properties of 3D LaCl and 2D GdCl are checked with different U, which are shown in
Fig. \ref{bands}(e), \ref{bands}(f), \ref{bands}(g) and \ref{bands}(h). By increasing U, the depth of the band
inversion at $\Gamma$ point decreases monotonously for both 3D LaCl and 2D GdCl. However, the nodal lines in 3D LaCl
and QAHI in 2D GdCl can still be found when $U= 8$ eV. Therefore, it is very promising to find these nontrivial
phases in ferromagnetic LnX.

\section{3D QAHI and Chern semimetal}

\begin{figure}[tp]
\centering
\includegraphics[width=2.8in]{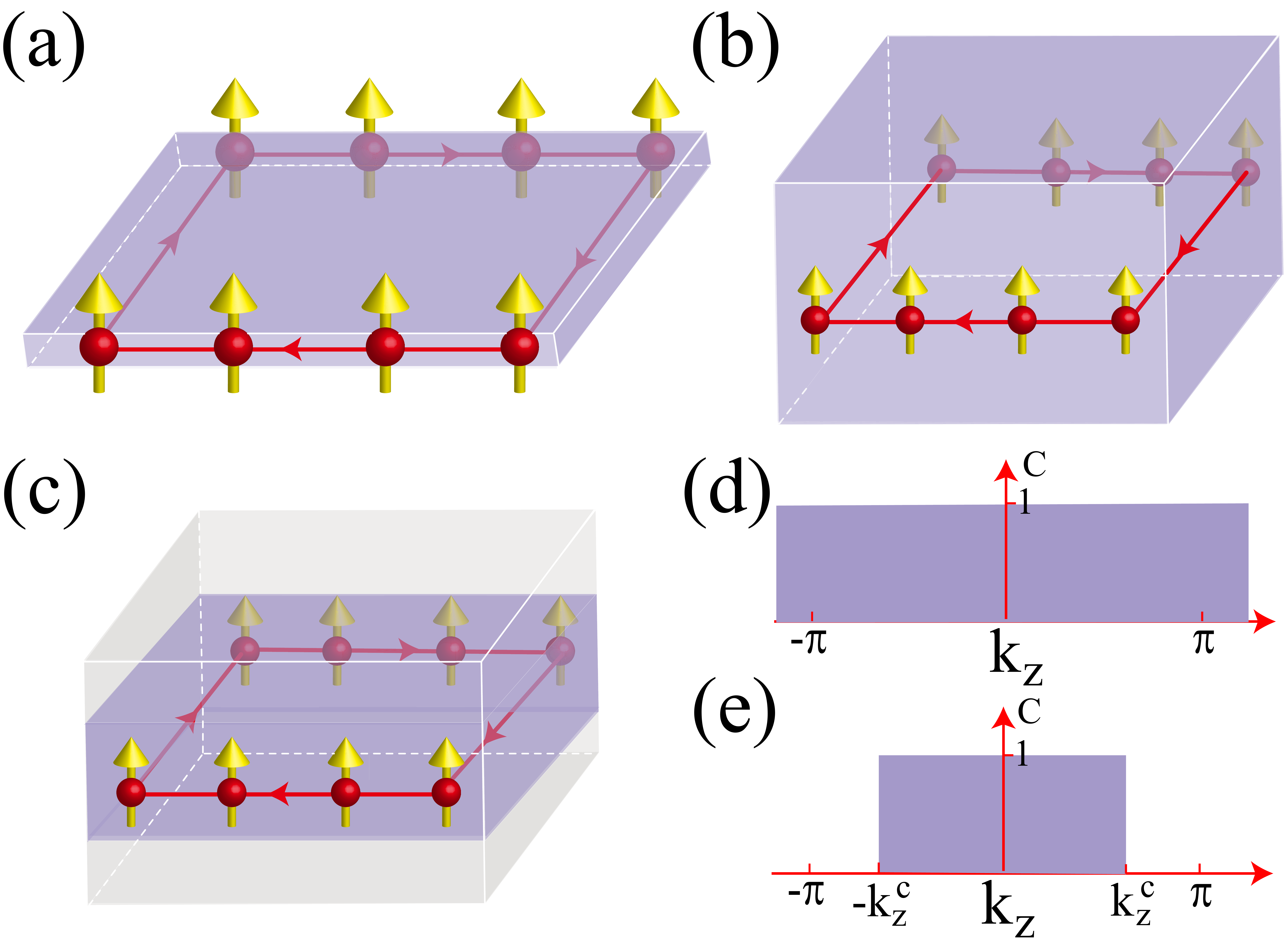}
      \caption{(Color online) (a, b, c) Schematics of 2D QAHI, 3D QAHI and Chern semimetal, respectively. (d, e) The Chern number C
      as function of $k_z$ for 3D QAHI and Chern semimetal, respectively.}
\label{qahi}
\end{figure}

2D QAHI  characterized by nonzero Chern number and quantized Hall conductance exhibits robust dissipationless chiral
edge states in the bulk band gap, as shown in Fig. \ref{qahi}(a). As discussed in the main text, the 3D material
constructed by stacking the 2D QAHI with C=1 may evolve into two different topological nontrivial phases, including
3D QAHI and Chern semimetal (also called Weyl semimetal), as shown in Fig. \ref{qahi}(b) and \ref{qahi}(c),
respectively. When the interlayer binding strength is very weak,
the 3D material is a 3D QAHI with $C=1$ for all $k_x$-, $k_y$- or $k_z$-fixed planes in the 3D BZ, which is consistent with
previous result\cite{halperin1987possible}. To be specific, we choose $k_z$-fixed planes as examples for the
illustration, as shown in Fig. \ref{qahi}(d). For 3D material with strong interlayer
binding strength (Chern semimetal), the Chern number equals to 1 for the plane with $-k_z^c < k_z < k_z^c $,
and equals to 0 for the plane with $ -\pi<k_z <- k_z^c$ or $k_z^c<k_z <\pi$, as shown in \ref{qahi}(e).
At the critical plane with $k_z=-k_z^c$ ($k_z=k_z^c$), there is a Weyl node with chiral charge $1$ ($-1$).
Although the Chern semimetal has been well studied\cite{xu2011chern}, the research on 3D QAHI is very few
owing to the lack of  promising candidates. Therefore, 3D GdX  without trivial pocket will greatly facilitate
the research on the nontrivial properties in 3D QAHI.

~\\



\begin{thebibliography}{70}%
\makeatletter
\providecommand \@ifxundefined [1]{%
 \@ifx{#1\undefined}
}%
\providecommand \@ifnum [1]{%
 \ifnum #1\expandafter \@firstoftwo
 \else \expandafter \@secondoftwo
 \fi
}%
\providecommand \@ifx [1]{%
 \ifx #1\expandafter \@firstoftwo
 \else \expandafter \@secondoftwo
 \fi
}%
\providecommand \natexlab [1]{#1}%
\providecommand \enquote  [1]{``#1''}%
\providecommand \bibnamefont  [1]{#1}%
\providecommand \bibfnamefont [1]{#1}%
\providecommand \citenamefont [1]{#1}%
\providecommand \href@noop [0]{\@secondoftwo}%
\providecommand \href [0]{\begingroup \@sanitize@url \@href}%
\providecommand \@href[1]{\@@startlink{#1}\@@href}%
\providecommand \@@href[1]{\endgroup#1\@@endlink}%
\providecommand \@sanitize@url [0]{\catcode `\\12\catcode `\$12\catcode
  `\&12\catcode `\#12\catcode `\^12\catcode `\_12\catcode `\%12\relax}%
\providecommand \@@startlink[1]{}%
\providecommand \@@endlink[0]{}%
\providecommand \url  [0]{\begingroup\@sanitize@url \@url }%
\providecommand \@url [1]{\endgroup\@href {#1}{\urlprefix }}%
\providecommand \urlprefix  [0]{URL }%
\providecommand \Eprint [0]{\href }%
\providecommand \doibase [0]{http://dx.doi.org/}%
\providecommand \selectlanguage [0]{\@gobble}%
\providecommand \bibinfo  [0]{\@secondoftwo}%
\providecommand \bibfield  [0]{\@secondoftwo}%
\providecommand \translation [1]{[#1]}%
\providecommand \BibitemOpen [0]{}%
\providecommand \bibitemStop [0]{}%
\providecommand \bibitemNoStop [0]{.\EOS\space}%
\providecommand \EOS [0]{\spacefactor3000\relax}%
\providecommand \BibitemShut  [1]{\csname bibitem#1\endcsname}%
\let\auto@bib@innerbib\@empty
\bibitem [{\citenamefont {Burkov}\ \emph {et~al.}(2011)\citenamefont {Burkov},
  \citenamefont {Hook},\ and\ \citenamefont {Balents}}]{burkov2011topological}%
  \BibitemOpen
  \bibfield  {author} {\bibinfo {author} {\bibfnamefont {A.}~\bibnamefont
  {Burkov}}, \bibinfo {author} {\bibfnamefont {M.}~\bibnamefont {Hook}}, \ and\
  \bibinfo {author} {\bibfnamefont {L.}~\bibnamefont {Balents}},\ }\href@noop
  {} {\bibfield  {journal} {\bibinfo  {journal} {Physical Review B}\ }\textbf
  {\bibinfo {volume} {84}},\ \bibinfo {pages} {235126} (\bibinfo {year}
  {2011})}\BibitemShut {NoStop}%
\bibitem [{\citenamefont {Fang}\ \emph {et~al.}(2015)\citenamefont {Fang},
  \citenamefont {Chen}, \citenamefont {Kee},\ and\ \citenamefont
  {Fu}}]{fang2015topological}%
  \BibitemOpen
  \bibfield  {author} {\bibinfo {author} {\bibfnamefont {C.}~\bibnamefont
  {Fang}}, \bibinfo {author} {\bibfnamefont {Y.}~\bibnamefont {Chen}}, \bibinfo
  {author} {\bibfnamefont {H.-Y.}\ \bibnamefont {Kee}}, \ and\ \bibinfo
  {author} {\bibfnamefont {L.}~\bibnamefont {Fu}},\ }\href@noop {} {\bibfield
  {journal} {\bibinfo  {journal} {Physical Review B}\ }\textbf {\bibinfo
  {volume} {92}},\ \bibinfo {pages} {081201} (\bibinfo {year}
  {2015})}\BibitemShut {NoStop}%
\bibitem [{\citenamefont {Weng}\ \emph {et~al.}(2016)\citenamefont {Weng},
  \citenamefont {Dai},\ and\ \citenamefont {Fang}}]{weng2016topological}%
  \BibitemOpen
  \bibfield  {author} {\bibinfo {author} {\bibfnamefont {H.}~\bibnamefont
  {Weng}}, \bibinfo {author} {\bibfnamefont {X.}~\bibnamefont {Dai}}, \ and\
  \bibinfo {author} {\bibfnamefont {Z.}~\bibnamefont {Fang}},\ }\href@noop {}
  {\bibfield  {journal} {\bibinfo  {journal} {J. Phys.: Condens. Matter}\
  }\textbf {\bibinfo {volume} {28}},\ \bibinfo {pages} {303001} (\bibinfo
  {year} {2016})}\BibitemShut {NoStop}%
\bibitem [{\citenamefont {Young}\ \emph {et~al.}(2012)\citenamefont {Young},
  \citenamefont {Zaheer}, \citenamefont {Teo}, \citenamefont {Kane},
  \citenamefont {Mele},\ and\ \citenamefont {Rappe}}]{young2012dirac}%
  \BibitemOpen
  \bibfield  {author} {\bibinfo {author} {\bibfnamefont {S.~M.}\ \bibnamefont
  {Young}}, \bibinfo {author} {\bibfnamefont {S.}~\bibnamefont {Zaheer}},
  \bibinfo {author} {\bibfnamefont {J.~C.}\ \bibnamefont {Teo}}, \bibinfo
  {author} {\bibfnamefont {C.~L.}\ \bibnamefont {Kane}}, \bibinfo {author}
  {\bibfnamefont {E.~J.}\ \bibnamefont {Mele}}, \ and\ \bibinfo {author}
  {\bibfnamefont {A.~M.}\ \bibnamefont {Rappe}},\ }\href@noop {} {\bibfield
  {journal} {\bibinfo  {journal} {Physical review letters}\ }\textbf {\bibinfo
  {volume} {108}},\ \bibinfo {pages} {140405} (\bibinfo {year}
  {2012})}\BibitemShut {NoStop}%
\bibitem [{\citenamefont {Wang}\ \emph {et~al.}(2012)\citenamefont {Wang},
  \citenamefont {Sun}, \citenamefont {Chen}, \citenamefont {Franchini},
  \citenamefont {Xu}, \citenamefont {Weng}, \citenamefont {Dai},\ and\
  \citenamefont {Fang}}]{wang2012dirac}%
  \BibitemOpen
  \bibfield  {author} {\bibinfo {author} {\bibfnamefont {Z.}~\bibnamefont
  {Wang}}, \bibinfo {author} {\bibfnamefont {Y.}~\bibnamefont {Sun}}, \bibinfo
  {author} {\bibfnamefont {X.-Q.}\ \bibnamefont {Chen}}, \bibinfo {author}
  {\bibfnamefont {C.}~\bibnamefont {Franchini}}, \bibinfo {author}
  {\bibfnamefont {G.}~\bibnamefont {Xu}}, \bibinfo {author} {\bibfnamefont
  {H.}~\bibnamefont {Weng}}, \bibinfo {author} {\bibfnamefont {X.}~\bibnamefont
  {Dai}}, \ and\ \bibinfo {author} {\bibfnamefont {Z.}~\bibnamefont {Fang}},\
  }\href@noop {} {\bibfield  {journal} {\bibinfo  {journal} {Physical Review
  B}\ }\textbf {\bibinfo {volume} {85}},\ \bibinfo {pages} {195320} (\bibinfo
  {year} {2012})}\BibitemShut {NoStop}%
\bibitem [{\citenamefont {Wang}\ \emph {et~al.}(2013)\citenamefont {Wang},
  \citenamefont {Weng}, \citenamefont {Wu}, \citenamefont {Dai},\ and\
  \citenamefont {Fang}}]{wang2013three}%
  \BibitemOpen
  \bibfield  {author} {\bibinfo {author} {\bibfnamefont {Z.}~\bibnamefont
  {Wang}}, \bibinfo {author} {\bibfnamefont {H.}~\bibnamefont {Weng}}, \bibinfo
  {author} {\bibfnamefont {Q.}~\bibnamefont {Wu}}, \bibinfo {author}
  {\bibfnamefont {X.}~\bibnamefont {Dai}}, \ and\ \bibinfo {author}
  {\bibfnamefont {Z.}~\bibnamefont {Fang}},\ }\href@noop {} {\bibfield
  {journal} {\bibinfo  {journal} {Physical Review B}\ }\textbf {\bibinfo
  {volume} {88}},\ \bibinfo {pages} {125427} (\bibinfo {year}
  {2013})}\BibitemShut {NoStop}%
\bibitem [{\citenamefont {Nielsen}\ and\ \citenamefont
  {Ninomiya}(1983)}]{nielsen1983adler}%
  \BibitemOpen
  \bibfield  {author} {\bibinfo {author} {\bibfnamefont {H.~B.}\ \bibnamefont
  {Nielsen}}\ and\ \bibinfo {author} {\bibfnamefont {M.}~\bibnamefont
  {Ninomiya}},\ }\href@noop {} {\bibfield  {journal} {\bibinfo  {journal}
  {Physics Letters B}\ }\textbf {\bibinfo {volume} {130}},\ \bibinfo {pages}
  {389} (\bibinfo {year} {1983})}\BibitemShut {NoStop}%
\bibitem [{\citenamefont {Wan}\ \emph {et~al.}(2011)\citenamefont {Wan},
  \citenamefont {Turner}, \citenamefont {Vishwanath},\ and\ \citenamefont
  {Savrasov}}]{wan2011topological}%
  \BibitemOpen
  \bibfield  {author} {\bibinfo {author} {\bibfnamefont {X.}~\bibnamefont
  {Wan}}, \bibinfo {author} {\bibfnamefont {A.~M.}\ \bibnamefont {Turner}},
  \bibinfo {author} {\bibfnamefont {A.}~\bibnamefont {Vishwanath}}, \ and\
  \bibinfo {author} {\bibfnamefont {S.~Y.}\ \bibnamefont {Savrasov}},\
  }\href@noop {} {\bibfield  {journal} {\bibinfo  {journal} {Physical Review
  B}\ }\textbf {\bibinfo {volume} {83}},\ \bibinfo {pages} {205101} (\bibinfo
  {year} {2011})}\BibitemShut {NoStop}%
\bibitem [{\citenamefont {Balents}(2011)}]{balents2011weyl}%
  \BibitemOpen
  \bibfield  {author} {\bibinfo {author} {\bibfnamefont {L.}~\bibnamefont
  {Balents}},\ }\href@noop {} {\bibfield  {journal} {\bibinfo  {journal}
  {Physics}\ }\textbf {\bibinfo {volume} {4}},\ \bibinfo {pages} {36} (\bibinfo
  {year} {2011})}\BibitemShut {NoStop}%
\bibitem [{\citenamefont {Yu}\ \emph {et~al.}(2017)\citenamefont {Yu},
  \citenamefont {Wu}, \citenamefont {Fang},\ and\ \citenamefont
  {Weng}}]{yuHfC}%
  \BibitemOpen
  \bibfield  {author} {\bibinfo {author} {\bibfnamefont {R.}~\bibnamefont
  {Yu}}, \bibinfo {author} {\bibfnamefont {Q.}~\bibnamefont {Wu}}, \bibinfo
  {author} {\bibfnamefont {Z.}~\bibnamefont {Fang}}, \ and\ \bibinfo {author}
  {\bibfnamefont {H.}~\bibnamefont {Weng}},\ }\href {\doibase
  10.1103/PhysRevLett.119.036401} {\bibfield  {journal} {\bibinfo  {journal}
  {Phys. Rev. Lett.}\ }\textbf {\bibinfo {volume} {119}},\ \bibinfo {pages}
  {036401} (\bibinfo {year} {2017})}\BibitemShut {NoStop}%
\bibitem [{\citenamefont {Du}\ \emph {et~al.}(2017)\citenamefont {Du},
  \citenamefont {Bo}, \citenamefont {Wang}, \citenamefont {Kan}, \citenamefont
  {Duan}, \citenamefont {Savrasov},\ and\ \citenamefont
  {Wan}}]{du2017emergence}%
  \BibitemOpen
  \bibfield  {author} {\bibinfo {author} {\bibfnamefont {Y.}~\bibnamefont
  {Du}}, \bibinfo {author} {\bibfnamefont {X.}~\bibnamefont {Bo}}, \bibinfo
  {author} {\bibfnamefont {D.}~\bibnamefont {Wang}}, \bibinfo {author}
  {\bibfnamefont {E.-j.}\ \bibnamefont {Kan}}, \bibinfo {author} {\bibfnamefont
  {C.-G.}\ \bibnamefont {Duan}}, \bibinfo {author} {\bibfnamefont {S.~Y.}\
  \bibnamefont {Savrasov}}, \ and\ \bibinfo {author} {\bibfnamefont
  {X.}~\bibnamefont {Wan}},\ }\href@noop {} {\bibfield  {journal} {\bibinfo
  {journal} {arXiv preprint arXiv:1708.04556}\ } (\bibinfo {year}
  {2017})}\BibitemShut {NoStop}%
\bibitem [{\citenamefont {Kim}\ \emph {et~al.}(2015)\citenamefont {Kim},
  \citenamefont {Wieder}, \citenamefont {Kane},\ and\ \citenamefont
  {Rappe}}]{kim2015dirac}%
  \BibitemOpen
  \bibfield  {author} {\bibinfo {author} {\bibfnamefont {Y.}~\bibnamefont
  {Kim}}, \bibinfo {author} {\bibfnamefont {B.~J.}\ \bibnamefont {Wieder}},
  \bibinfo {author} {\bibfnamefont {C.}~\bibnamefont {Kane}}, \ and\ \bibinfo
  {author} {\bibfnamefont {A.~M.}\ \bibnamefont {Rappe}},\ }\href@noop {}
  {\bibfield  {journal} {\bibinfo  {journal} {Physical review letters}\
  }\textbf {\bibinfo {volume} {115}},\ \bibinfo {pages} {036806} (\bibinfo
  {year} {2015})}\BibitemShut {NoStop}%
\bibitem [{\citenamefont {Yu}\ \emph {et~al.}(2015)\citenamefont {Yu},
  \citenamefont {Weng}, \citenamefont {Fang}, \citenamefont {Dai},\ and\
  \citenamefont {Hu}}]{yu2015topological}%
  \BibitemOpen
  \bibfield  {author} {\bibinfo {author} {\bibfnamefont {R.}~\bibnamefont
  {Yu}}, \bibinfo {author} {\bibfnamefont {H.}~\bibnamefont {Weng}}, \bibinfo
  {author} {\bibfnamefont {Z.}~\bibnamefont {Fang}}, \bibinfo {author}
  {\bibfnamefont {X.}~\bibnamefont {Dai}}, \ and\ \bibinfo {author}
  {\bibfnamefont {X.}~\bibnamefont {Hu}},\ }\href@noop {} {\bibfield  {journal}
  {\bibinfo  {journal} {Physical review letters}\ }\textbf {\bibinfo {volume}
  {115}},\ \bibinfo {pages} {036807} (\bibinfo {year} {2015})}\BibitemShut
  {NoStop}%
\bibitem [{\citenamefont {Carter}\ \emph {et~al.}(2012)\citenamefont {Carter},
  \citenamefont {Shankar}, \citenamefont {Zeb},\ and\ \citenamefont
  {Kee}}]{srIo3}%
  \BibitemOpen
  \bibfield  {author} {\bibinfo {author} {\bibfnamefont {J.-M.}\ \bibnamefont
  {Carter}}, \bibinfo {author} {\bibfnamefont {V.~V.}\ \bibnamefont {Shankar}},
  \bibinfo {author} {\bibfnamefont {M.~A.}\ \bibnamefont {Zeb}}, \ and\
  \bibinfo {author} {\bibfnamefont {H.-Y.}\ \bibnamefont {Kee}},\ }\href
  {\doibase 10.1103/PhysRevB.85.115105} {\bibfield  {journal} {\bibinfo
  {journal} {Phys. Rev. B}\ }\textbf {\bibinfo {volume} {85}},\ \bibinfo
  {pages} {115105} (\bibinfo {year} {2012})}\BibitemShut {NoStop}%
\bibitem [{\citenamefont {Nie}\ \emph {et~al.}(2017)\citenamefont {Nie},
  \citenamefont {Xu}, \citenamefont {Prinz},\ and\ \citenamefont
  {Zhang}}]{Nie03102017}%
  \BibitemOpen
  \bibfield  {author} {\bibinfo {author} {\bibfnamefont {S.}~\bibnamefont
  {Nie}}, \bibinfo {author} {\bibfnamefont {G.}~\bibnamefont {Xu}}, \bibinfo
  {author} {\bibfnamefont {F.~B.}\ \bibnamefont {Prinz}}, \ and\ \bibinfo
  {author} {\bibfnamefont {S.-c.}\ \bibnamefont {Zhang}},\ }\href {\doibase
  10.1073/pnas.1713261114} {\bibfield  {journal} {\bibinfo  {journal}
  {Proceedings of the National Academy of Sciences}\ }\textbf {\bibinfo
  {volume} {114}},\ \bibinfo {pages} {10596} (\bibinfo {year}
  {2017})}\BibitemShut {NoStop}%
\bibitem [{\citenamefont {Dolui}\ \emph {et~al.}(2015)\citenamefont {Dolui},
  \citenamefont {Ray},\ and\ \citenamefont {Das}}]{LaXqahe}%
  \BibitemOpen
  \bibfield  {author} {\bibinfo {author} {\bibfnamefont {K.}~\bibnamefont
  {Dolui}}, \bibinfo {author} {\bibfnamefont {S.}~\bibnamefont {Ray}}, \ and\
  \bibinfo {author} {\bibfnamefont {T.}~\bibnamefont {Das}},\ }\href {\doibase
  10.1103/PhysRevB.92.205133} {\bibfield  {journal} {\bibinfo  {journal} {Phys.
  Rev. B}\ }\textbf {\bibinfo {volume} {92}},\ \bibinfo {pages} {205133}
  (\bibinfo {year} {2015})}\BibitemShut {NoStop}%
\bibitem [{\citenamefont {Wu}(2017)}]{wu2017high}%
  \BibitemOpen
  \bibfield  {author} {\bibinfo {author} {\bibfnamefont {M.}~\bibnamefont
  {Wu}},\ }\href@noop {} {\bibfield  {journal} {\bibinfo  {journal} {2D
  Materials}\ }\textbf {\bibinfo {volume} {4}},\ \bibinfo {pages} {021014}
  (\bibinfo {year} {2017})}\BibitemShut {NoStop}%
\bibitem [{\citenamefont {Wang}\ \emph {et~al.}(2016)\citenamefont {Wang},
  \citenamefont {Weng}, \citenamefont {Nie}, \citenamefont {Fang},
  \citenamefont {Kawazoe},\ and\ \citenamefont {Chen}}]{wang2016body}%
  \BibitemOpen
  \bibfield  {author} {\bibinfo {author} {\bibfnamefont {J.-T.}\ \bibnamefont
  {Wang}}, \bibinfo {author} {\bibfnamefont {H.}~\bibnamefont {Weng}}, \bibinfo
  {author} {\bibfnamefont {S.}~\bibnamefont {Nie}}, \bibinfo {author}
  {\bibfnamefont {Z.}~\bibnamefont {Fang}}, \bibinfo {author} {\bibfnamefont
  {Y.}~\bibnamefont {Kawazoe}}, \ and\ \bibinfo {author} {\bibfnamefont
  {C.}~\bibnamefont {Chen}},\ }\href@noop {} {\bibfield  {journal} {\bibinfo
  {journal} {Physical review letters}\ }\textbf {\bibinfo {volume} {116}},\
  \bibinfo {pages} {195501} (\bibinfo {year} {2016})}\BibitemShut {NoStop}%
\bibitem [{\citenamefont {Chen}\ \emph
  {et~al.}(2015{\natexlab{a}})\citenamefont {Chen}, \citenamefont {Xie},
  \citenamefont {Yang}, \citenamefont {Pan}, \citenamefont {Zhang},
  \citenamefont {Cohen},\ and\ \citenamefont {Zhang}}]{chen2015nanostructured}%
  \BibitemOpen
  \bibfield  {author} {\bibinfo {author} {\bibfnamefont {Y.}~\bibnamefont
  {Chen}}, \bibinfo {author} {\bibfnamefont {Y.}~\bibnamefont {Xie}}, \bibinfo
  {author} {\bibfnamefont {S.~A.}\ \bibnamefont {Yang}}, \bibinfo {author}
  {\bibfnamefont {H.}~\bibnamefont {Pan}}, \bibinfo {author} {\bibfnamefont
  {F.}~\bibnamefont {Zhang}}, \bibinfo {author} {\bibfnamefont {M.~L.}\
  \bibnamefont {Cohen}}, \ and\ \bibinfo {author} {\bibfnamefont
  {S.}~\bibnamefont {Zhang}},\ }\href@noop {} {\bibfield  {journal} {\bibinfo
  {journal} {Nano letters}\ }\textbf {\bibinfo {volume} {15}},\ \bibinfo
  {pages} {6974} (\bibinfo {year} {2015}{\natexlab{a}})}\BibitemShut {NoStop}%
\bibitem [{\citenamefont {Bzdu{\v{s}}ek}\ \emph {et~al.}(2016)\citenamefont
  {Bzdu{\v{s}}ek}, \citenamefont {Wu}, \citenamefont {R{\"u}egg}, \citenamefont
  {Sigrist},\ and\ \citenamefont {Soluyanov}}]{bzduvsek2016nodal}%
  \BibitemOpen
  \bibfield  {author} {\bibinfo {author} {\bibfnamefont {T.}~\bibnamefont
  {Bzdu{\v{s}}ek}}, \bibinfo {author} {\bibfnamefont {Q.}~\bibnamefont {Wu}},
  \bibinfo {author} {\bibfnamefont {A.}~\bibnamefont {R{\"u}egg}}, \bibinfo
  {author} {\bibfnamefont {M.}~\bibnamefont {Sigrist}}, \ and\ \bibinfo
  {author} {\bibfnamefont {A.~A.}\ \bibnamefont {Soluyanov}},\ }\href@noop {}
  {\bibfield  {journal} {\bibinfo  {journal} {Nature}\ }\textbf {\bibinfo
  {volume} {538}},\ \bibinfo {pages} {75} (\bibinfo {year} {2016})}\BibitemShut
  {NoStop}%
\bibitem [{\citenamefont {Kopnin}\ \emph {et~al.}(2011)\citenamefont {Kopnin},
  \citenamefont {Heikkil\"a},\ and\ \citenamefont {Volovik}}]{hsp1}%
  \BibitemOpen
  \bibfield  {author} {\bibinfo {author} {\bibfnamefont {N.~B.}\ \bibnamefont
  {Kopnin}}, \bibinfo {author} {\bibfnamefont {T.~T.}\ \bibnamefont
  {Heikkil\"a}}, \ and\ \bibinfo {author} {\bibfnamefont {G.~E.}\ \bibnamefont
  {Volovik}},\ }\href {\doibase 10.1103/PhysRevB.83.220503} {\bibfield
  {journal} {\bibinfo  {journal} {Phys. Rev. B}\ }\textbf {\bibinfo {volume}
  {83}},\ \bibinfo {pages} {220503} (\bibinfo {year} {2011})}\BibitemShut
  {NoStop}%
\bibitem [{\citenamefont {Volovik}(2015)}]{volovik2015standard}%
  \BibitemOpen
  \bibfield  {author} {\bibinfo {author} {\bibfnamefont {G.}~\bibnamefont
  {Volovik}},\ }\href@noop {} {\bibfield  {journal} {\bibinfo  {journal}
  {Physica Scripta}\ }\textbf {\bibinfo {volume} {2015}},\ \bibinfo {pages}
  {014014} (\bibinfo {year} {2015})}\BibitemShut {NoStop}%
\bibitem [{\citenamefont {Heikkil{\"a}}\ and\ \citenamefont
  {Volovik}(2016)}]{heikkila2016flat}%
  \BibitemOpen
  \bibfield  {author} {\bibinfo {author} {\bibfnamefont {T.~T.}\ \bibnamefont
  {Heikkil{\"a}}}\ and\ \bibinfo {author} {\bibfnamefont {G.~E.}\ \bibnamefont
  {Volovik}},\ }in\ \href@noop {} {\emph {\bibinfo {booktitle} {Basic Physics
  of Functionalized Graphite}}}\ (\bibinfo  {publisher} {Springer},\ \bibinfo
  {year} {2016})\ pp.\ \bibinfo {pages} {123--143}\BibitemShut {NoStop}%
\bibitem [{\citenamefont {Rhim}\ and\ \citenamefont
  {Kim}(2015)}]{rhim2015landau}%
  \BibitemOpen
  \bibfield  {author} {\bibinfo {author} {\bibfnamefont {J.-W.}\ \bibnamefont
  {Rhim}}\ and\ \bibinfo {author} {\bibfnamefont {Y.~B.}\ \bibnamefont {Kim}},\
  }\href@noop {} {\bibfield  {journal} {\bibinfo  {journal} {Physical Review
  B}\ }\textbf {\bibinfo {volume} {92}},\ \bibinfo {pages} {045126} (\bibinfo
  {year} {2015})}\BibitemShut {NoStop}%
\bibitem [{\citenamefont {Huh}\ \emph {et~al.}(2016)\citenamefont {Huh},
  \citenamefont {Moon},\ and\ \citenamefont {Kim}}]{huh2016long}%
  \BibitemOpen
  \bibfield  {author} {\bibinfo {author} {\bibfnamefont {Y.}~\bibnamefont
  {Huh}}, \bibinfo {author} {\bibfnamefont {E.-G.}\ \bibnamefont {Moon}}, \
  and\ \bibinfo {author} {\bibfnamefont {Y.~B.}\ \bibnamefont {Kim}},\
  }\href@noop {} {\bibfield  {journal} {\bibinfo  {journal} {Physical Review
  B}\ }\textbf {\bibinfo {volume} {93}},\ \bibinfo {pages} {035138} (\bibinfo
  {year} {2016})}\BibitemShut {NoStop}%
\bibitem [{\citenamefont {Weng}\ \emph {et~al.}(2015)\citenamefont {Weng},
  \citenamefont {Liang}, \citenamefont {Xu}, \citenamefont {Yu}, \citenamefont
  {Fang}, \citenamefont {Dai},\ and\ \citenamefont {Kawazoe}}]{wengC}%
  \BibitemOpen
  \bibfield  {author} {\bibinfo {author} {\bibfnamefont {H.}~\bibnamefont
  {Weng}}, \bibinfo {author} {\bibfnamefont {Y.}~\bibnamefont {Liang}},
  \bibinfo {author} {\bibfnamefont {Q.}~\bibnamefont {Xu}}, \bibinfo {author}
  {\bibfnamefont {R.}~\bibnamefont {Yu}}, \bibinfo {author} {\bibfnamefont
  {Z.}~\bibnamefont {Fang}}, \bibinfo {author} {\bibfnamefont {X.}~\bibnamefont
  {Dai}}, \ and\ \bibinfo {author} {\bibfnamefont {Y.}~\bibnamefont
  {Kawazoe}},\ }\href {\doibase 10.1103/PhysRevB.92.045108} {\bibfield
  {journal} {\bibinfo  {journal} {Phys. Rev. B}\ }\textbf {\bibinfo {volume}
  {92}},\ \bibinfo {pages} {045108} (\bibinfo {year} {2015})}\BibitemShut
  {NoStop}%
\bibitem [{\citenamefont {Xie}\ \emph {et~al.}(2015)\citenamefont {Xie},
  \citenamefont {Schoop}, \citenamefont {Seibel}, \citenamefont {Gibson},
  \citenamefont {Xie},\ and\ \citenamefont {Cava}}]{xie2015new}%
  \BibitemOpen
  \bibfield  {author} {\bibinfo {author} {\bibfnamefont {L.~S.}\ \bibnamefont
  {Xie}}, \bibinfo {author} {\bibfnamefont {L.~M.}\ \bibnamefont {Schoop}},
  \bibinfo {author} {\bibfnamefont {E.~M.}\ \bibnamefont {Seibel}}, \bibinfo
  {author} {\bibfnamefont {Q.~D.}\ \bibnamefont {Gibson}}, \bibinfo {author}
  {\bibfnamefont {W.}~\bibnamefont {Xie}}, \ and\ \bibinfo {author}
  {\bibfnamefont {R.~J.}\ \bibnamefont {Cava}},\ }\href@noop {} {\bibfield
  {journal} {\bibinfo  {journal} {Apl Materials}\ }\textbf {\bibinfo {volume}
  {3}},\ \bibinfo {pages} {083602} (\bibinfo {year} {2015})}\BibitemShut
  {NoStop}%
\bibitem [{\citenamefont {Chan}\ \emph {et~al.}(2016)\citenamefont {Chan},
  \citenamefont {Chiu}, \citenamefont {Chou},\ and\ \citenamefont
  {Schnyder}}]{chan2016}%
  \BibitemOpen
  \bibfield  {author} {\bibinfo {author} {\bibfnamefont {Y.-H.}\ \bibnamefont
  {Chan}}, \bibinfo {author} {\bibfnamefont {C.-K.}\ \bibnamefont {Chiu}},
  \bibinfo {author} {\bibfnamefont {M.}~\bibnamefont {Chou}}, \ and\ \bibinfo
  {author} {\bibfnamefont {A.~P.}\ \bibnamefont {Schnyder}},\ }\href@noop {}
  {\bibfield  {journal} {\bibinfo  {journal} {Physical Review B}\ }\textbf
  {\bibinfo {volume} {93}},\ \bibinfo {pages} {205132} (\bibinfo {year}
  {2016})}\BibitemShut {NoStop}%
\bibitem [{\citenamefont {Hirayama}\ \emph {et~al.}(2017)\citenamefont
  {Hirayama}, \citenamefont {Okugawa}, \citenamefont {Miyake},\ and\
  \citenamefont {Murakami}}]{hirayama2017topological}%
  \BibitemOpen
  \bibfield  {author} {\bibinfo {author} {\bibfnamefont {M.}~\bibnamefont
  {Hirayama}}, \bibinfo {author} {\bibfnamefont {R.}~\bibnamefont {Okugawa}},
  \bibinfo {author} {\bibfnamefont {T.}~\bibnamefont {Miyake}}, \ and\ \bibinfo
  {author} {\bibfnamefont {S.}~\bibnamefont {Murakami}},\ }\href@noop {}
  {\bibfield  {journal} {\bibinfo  {journal} {Nature communications}\ }\textbf
  {\bibinfo {volume} {8}} (\bibinfo {year} {2017})}\BibitemShut {NoStop}%
\bibitem [{\citenamefont {Zhao}\ \emph {et~al.}(2016)\citenamefont {Zhao},
  \citenamefont {Yu}, \citenamefont {Weng},\ and\ \citenamefont
  {Fang}}]{jzhoublack}%
  \BibitemOpen
  \bibfield  {author} {\bibinfo {author} {\bibfnamefont {J.}~\bibnamefont
  {Zhao}}, \bibinfo {author} {\bibfnamefont {R.}~\bibnamefont {Yu}}, \bibinfo
  {author} {\bibfnamefont {H.}~\bibnamefont {Weng}}, \ and\ \bibinfo {author}
  {\bibfnamefont {Z.}~\bibnamefont {Fang}},\ }\href {\doibase
  10.1103/PhysRevB.94.195104} {\bibfield  {journal} {\bibinfo  {journal} {Phys.
  Rev. B}\ }\textbf {\bibinfo {volume} {94}},\ \bibinfo {pages} {195104}
  (\bibinfo {year} {2016})}\BibitemShut {NoStop}%
\bibitem [{\citenamefont {Xu}\ \emph {et~al.}(2017)\citenamefont {Xu},
  \citenamefont {Yu}, \citenamefont {Fang}, \citenamefont {Dai},\ and\
  \citenamefont {Weng}}]{XuCaP3}%
  \BibitemOpen
  \bibfield  {author} {\bibinfo {author} {\bibfnamefont {Q.}~\bibnamefont
  {Xu}}, \bibinfo {author} {\bibfnamefont {R.}~\bibnamefont {Yu}}, \bibinfo
  {author} {\bibfnamefont {Z.}~\bibnamefont {Fang}}, \bibinfo {author}
  {\bibfnamefont {X.}~\bibnamefont {Dai}}, \ and\ \bibinfo {author}
  {\bibfnamefont {H.}~\bibnamefont {Weng}},\ }\href {\doibase
  10.1103/PhysRevB.95.045136} {\bibfield  {journal} {\bibinfo  {journal} {Phys.
  Rev. B}\ }\textbf {\bibinfo {volume} {95}},\ \bibinfo {pages} {045136}
  (\bibinfo {year} {2017})}\BibitemShut {NoStop}%
\bibitem [{\citenamefont {Chang}\ \emph {et~al.}(2017)\citenamefont {Chang},
  \citenamefont {Pletikosic}, \citenamefont {Kong}, \citenamefont {Bian},
  \citenamefont {Huang}, \citenamefont {Denlinger}, \citenamefont {Kushwaha},
  \citenamefont {Sinkovic}, \citenamefont {Jeng}, \citenamefont {Valla},
  \citenamefont {Xie},\ and\ \citenamefont {Cava}}]{chang2017realization}%
  \BibitemOpen
  \bibfield  {author} {\bibinfo {author} {\bibfnamefont {T.-R.}\ \bibnamefont
  {Chang}}, \bibinfo {author} {\bibfnamefont {I.}~\bibnamefont {Pletikosic}},
  \bibinfo {author} {\bibfnamefont {T.}~\bibnamefont {Kong}}, \bibinfo {author}
  {\bibfnamefont {G.}~\bibnamefont {Bian}}, \bibinfo {author} {\bibfnamefont
  {A.}~\bibnamefont {Huang}}, \bibinfo {author} {\bibfnamefont
  {J.}~\bibnamefont {Denlinger}}, \bibinfo {author} {\bibfnamefont {S.~K.}\
  \bibnamefont {Kushwaha}}, \bibinfo {author} {\bibfnamefont {B.}~\bibnamefont
  {Sinkovic}}, \bibinfo {author} {\bibfnamefont {H.-T.}\ \bibnamefont {Jeng}},
  \bibinfo {author} {\bibfnamefont {T.}~\bibnamefont {Valla}}, \bibinfo
  {author} {\bibfnamefont {W.}~\bibnamefont {Xie}}, \ and\ \bibinfo {author}
  {\bibfnamefont {R.~J.}\ \bibnamefont {Cava}},\ }\href@noop {} {\bibfield
  {journal} {\bibinfo  {journal} {arXiv preprint arXiv:1711.09167}\ } (\bibinfo
  {year} {2017})}\BibitemShut {NoStop}%
\bibitem [{\citenamefont {Feng}\ \emph {et~al.}(2018)\citenamefont {Feng},
  \citenamefont {Yue}, \citenamefont {Song}, \citenamefont {Wu},\ and\
  \citenamefont {Wen}}]{feng2018topological}%
  \BibitemOpen
  \bibfield  {author} {\bibinfo {author} {\bibfnamefont {X.}~\bibnamefont
  {Feng}}, \bibinfo {author} {\bibfnamefont {C.}~\bibnamefont {Yue}}, \bibinfo
  {author} {\bibfnamefont {Z.}~\bibnamefont {Song}}, \bibinfo {author}
  {\bibfnamefont {Q.}~\bibnamefont {Wu}}, \ and\ \bibinfo {author}
  {\bibfnamefont {B.}~\bibnamefont {Wen}},\ }\href@noop {} {\bibfield
  {journal} {\bibinfo  {journal} {Physical Review Materials}\ }\textbf
  {\bibinfo {volume} {2}},\ \bibinfo {pages} {014202} (\bibinfo {year}
  {2018})}\BibitemShut {NoStop}%
\bibitem [{\citenamefont {Wang}\ \emph {et~al.}(2018)\citenamefont {Wang},
  \citenamefont {Nie}, \citenamefont {Weng}, \citenamefont {Kawazoe},\ and\
  \citenamefont {Chen}}]{wang2018topological}%
  \BibitemOpen
  \bibfield  {author} {\bibinfo {author} {\bibfnamefont {J.-T.}\ \bibnamefont
  {Wang}}, \bibinfo {author} {\bibfnamefont {S.}~\bibnamefont {Nie}}, \bibinfo
  {author} {\bibfnamefont {H.}~\bibnamefont {Weng}}, \bibinfo {author}
  {\bibfnamefont {Y.}~\bibnamefont {Kawazoe}}, \ and\ \bibinfo {author}
  {\bibfnamefont {C.}~\bibnamefont {Chen}},\ }\href@noop {} {\bibfield
  {journal} {\bibinfo  {journal} {Physical Review Letters}\ }\textbf {\bibinfo
  {volume} {120}},\ \bibinfo {pages} {026402} (\bibinfo {year}
  {2018})}\BibitemShut {NoStop}%
\bibitem [{\citenamefont {Li}\ \emph {et~al.}(2016)\citenamefont {Li},
  \citenamefont {Ma}, \citenamefont {Cheng}, \citenamefont {Wang},
  \citenamefont {Li}, \citenamefont {Zhang}, \citenamefont {Li},\ and\
  \citenamefont {Chen}}]{li2016dirac}%
  \BibitemOpen
  \bibfield  {author} {\bibinfo {author} {\bibfnamefont {R.}~\bibnamefont
  {Li}}, \bibinfo {author} {\bibfnamefont {H.}~\bibnamefont {Ma}}, \bibinfo
  {author} {\bibfnamefont {X.}~\bibnamefont {Cheng}}, \bibinfo {author}
  {\bibfnamefont {S.}~\bibnamefont {Wang}}, \bibinfo {author} {\bibfnamefont
  {D.}~\bibnamefont {Li}}, \bibinfo {author} {\bibfnamefont {Z.}~\bibnamefont
  {Zhang}}, \bibinfo {author} {\bibfnamefont {Y.}~\bibnamefont {Li}}, \ and\
  \bibinfo {author} {\bibfnamefont {X.-Q.}\ \bibnamefont {Chen}},\ }\href@noop
  {} {\bibfield  {journal} {\bibinfo  {journal} {Physical review letters}\
  }\textbf {\bibinfo {volume} {117}},\ \bibinfo {pages} {096401} (\bibinfo
  {year} {2016})}\BibitemShut {NoStop}%
\bibitem [{\citenamefont {Chen}\ \emph
  {et~al.}(2015{\natexlab{b}})\citenamefont {Chen}, \citenamefont {Lu},\ and\
  \citenamefont {Kee}}]{chen2015topological}%
  \BibitemOpen
  \bibfield  {author} {\bibinfo {author} {\bibfnamefont {Y.}~\bibnamefont
  {Chen}}, \bibinfo {author} {\bibfnamefont {Y.-M.}\ \bibnamefont {Lu}}, \ and\
  \bibinfo {author} {\bibfnamefont {H.-Y.}\ \bibnamefont {Kee}},\ }\href@noop
  {} {\bibfield  {journal} {\bibinfo  {journal} {Nature communications}\
  }\textbf {\bibinfo {volume} {6}} (\bibinfo {year}
  {2015}{\natexlab{b}})}\BibitemShut {NoStop}%
\bibitem [{\citenamefont {Bian}\ \emph {et~al.}(2016)\citenamefont {Bian},
  \citenamefont {Chang}, \citenamefont {Sankar}, \citenamefont {Xu},
  \citenamefont {Zheng}, \citenamefont {Neupert}, \citenamefont {Chiu},
  \citenamefont {Huang}, \citenamefont {Chang}, \citenamefont {Belopolski},
  \citenamefont {Sanchez}, \citenamefont {Neupane}, \citenamefont {Alidoust},
  \citenamefont {Liu}, \citenamefont {Wang}, \citenamefont {Lee}, \citenamefont
  {Jeng}, \citenamefont {Zhang}, \citenamefont {Yuan}, \citenamefont {Jia},
  \citenamefont {Bansil}, \citenamefont {Chou}, \citenamefont {Lin},\ and\
  \citenamefont {Hasan}}]{bian2016topological}%
  \BibitemOpen
  \bibfield  {author} {\bibinfo {author} {\bibfnamefont {G.}~\bibnamefont
  {Bian}}, \bibinfo {author} {\bibfnamefont {T.-R.}\ \bibnamefont {Chang}},
  \bibinfo {author} {\bibfnamefont {R.}~\bibnamefont {Sankar}}, \bibinfo
  {author} {\bibfnamefont {S.-Y.}\ \bibnamefont {Xu}}, \bibinfo {author}
  {\bibfnamefont {H.}~\bibnamefont {Zheng}}, \bibinfo {author} {\bibfnamefont
  {T.}~\bibnamefont {Neupert}}, \bibinfo {author} {\bibfnamefont {C.-K.}\
  \bibnamefont {Chiu}}, \bibinfo {author} {\bibfnamefont {S.-M.}\ \bibnamefont
  {Huang}}, \bibinfo {author} {\bibfnamefont {G.}~\bibnamefont {Chang}},
  \bibinfo {author} {\bibfnamefont {I.}~\bibnamefont {Belopolski}}, \bibinfo
  {author} {\bibfnamefont {D.~S.}\ \bibnamefont {Sanchez}}, \bibinfo {author}
  {\bibfnamefont {M.}~\bibnamefont {Neupane}}, \bibinfo {author} {\bibfnamefont
  {N.}~\bibnamefont {Alidoust}}, \bibinfo {author} {\bibfnamefont
  {C.}~\bibnamefont {Liu}}, \bibinfo {author} {\bibfnamefont {B.}~\bibnamefont
  {Wang}}, \bibinfo {author} {\bibfnamefont {C.-C.}\ \bibnamefont {Lee}},
  \bibinfo {author} {\bibfnamefont {H.-T.}\ \bibnamefont {Jeng}}, \bibinfo
  {author} {\bibfnamefont {C.}~\bibnamefont {Zhang}}, \bibinfo {author}
  {\bibfnamefont {Z.}~\bibnamefont {Yuan}}, \bibinfo {author} {\bibfnamefont
  {S.}~\bibnamefont {Jia}}, \bibinfo {author} {\bibfnamefont {A.}~\bibnamefont
  {Bansil}}, \bibinfo {author} {\bibfnamefont {F.}~\bibnamefont {Chou}},
  \bibinfo {author} {\bibfnamefont {H.}~\bibnamefont {Lin}}, \ and\ \bibinfo
  {author} {\bibfnamefont {M.~Z.}\ \bibnamefont {Hasan}},\ }\href@noop {}
  {\bibfield  {journal} {\bibinfo  {journal} {Nature communications}\ }\textbf
  {\bibinfo {volume} {7}},\ \bibinfo {pages} {10556} (\bibinfo {year}
  {2016})}\BibitemShut {NoStop}%
\bibitem [{\citenamefont {Schoop}\ \emph {et~al.}(2016)\citenamefont {Schoop},
  \citenamefont {Ali}, \citenamefont {Stra{\ss}er}, \citenamefont {Topp},
  \citenamefont {Varykhalov}, \citenamefont {Marchenko}, \citenamefont
  {Duppel}, \citenamefont {Parkin}, \citenamefont {Lotsch},\ and\ \citenamefont
  {Ast}}]{schoop2016dirac}%
  \BibitemOpen
  \bibfield  {author} {\bibinfo {author} {\bibfnamefont {L.~M.}\ \bibnamefont
  {Schoop}}, \bibinfo {author} {\bibfnamefont {M.~N.}\ \bibnamefont {Ali}},
  \bibinfo {author} {\bibfnamefont {C.}~\bibnamefont {Stra{\ss}er}}, \bibinfo
  {author} {\bibfnamefont {A.}~\bibnamefont {Topp}}, \bibinfo {author}
  {\bibfnamefont {A.}~\bibnamefont {Varykhalov}}, \bibinfo {author}
  {\bibfnamefont {D.}~\bibnamefont {Marchenko}}, \bibinfo {author}
  {\bibfnamefont {V.}~\bibnamefont {Duppel}}, \bibinfo {author} {\bibfnamefont
  {S.~S.}\ \bibnamefont {Parkin}}, \bibinfo {author} {\bibfnamefont {B.~V.}\
  \bibnamefont {Lotsch}}, \ and\ \bibinfo {author} {\bibfnamefont {C.~R.}\
  \bibnamefont {Ast}},\ }\href@noop {} {\bibfield  {journal} {\bibinfo
  {journal} {Nature communications}\ }\textbf {\bibinfo {volume} {7}} (\bibinfo
  {year} {2016})}\BibitemShut {NoStop}%
\bibitem [{\citenamefont {Fu}\ \emph {et~al.}(2017)\citenamefont {Fu},
  \citenamefont {Yi}, \citenamefont {Zhang}, \citenamefont {Caputo},
  \citenamefont {Ma}, \citenamefont {Gao}, \citenamefont {Lv}, \citenamefont
  {Kong}, \citenamefont {Huang}, \citenamefont {Shi}, \citenamefont {Vladimir},
  \citenamefont {Fang}, \citenamefont {Weng}, \citenamefont {Shi},
  \citenamefont {Qian},\ and\ \citenamefont {Ding}}]{fu2017observation}%
  \BibitemOpen
  \bibfield  {author} {\bibinfo {author} {\bibfnamefont {B.}~\bibnamefont
  {Fu}}, \bibinfo {author} {\bibfnamefont {C.}~\bibnamefont {Yi}}, \bibinfo
  {author} {\bibfnamefont {T.}~\bibnamefont {Zhang}}, \bibinfo {author}
  {\bibfnamefont {M.}~\bibnamefont {Caputo}}, \bibinfo {author} {\bibfnamefont
  {J.}~\bibnamefont {Ma}}, \bibinfo {author} {\bibfnamefont {X.}~\bibnamefont
  {Gao}}, \bibinfo {author} {\bibfnamefont {B.}~\bibnamefont {Lv}}, \bibinfo
  {author} {\bibfnamefont {L.}~\bibnamefont {Kong}}, \bibinfo {author}
  {\bibfnamefont {Y.}~\bibnamefont {Huang}}, \bibinfo {author} {\bibfnamefont
  {M.}~\bibnamefont {Shi}}, \bibinfo {author} {\bibfnamefont {S.}~\bibnamefont
  {Vladimir}}, \bibinfo {author} {\bibfnamefont {C.}~\bibnamefont {Fang}},
  \bibinfo {author} {\bibfnamefont {H.}~\bibnamefont {Weng}}, \bibinfo {author}
  {\bibfnamefont {Y.}~\bibnamefont {Shi}}, \bibinfo {author} {\bibfnamefont
  {T.}~\bibnamefont {Qian}}, \ and\ \bibinfo {author} {\bibfnamefont
  {H.}~\bibnamefont {Ding}},\ }\href@noop {} {\bibfield  {journal} {\bibinfo
  {journal} {arXiv preprint arXiv:1712.00782}\ } (\bibinfo {year}
  {2017})}\BibitemShut {NoStop}%
\bibitem [{\citenamefont {Yu}\ \emph {et~al.}(2010)\citenamefont {Yu},
  \citenamefont {Zhang}, \citenamefont {Zhang}, \citenamefont {Zhang},
  \citenamefont {Dai},\ and\ \citenamefont {Fang}}]{yu2010quantized}%
  \BibitemOpen
  \bibfield  {author} {\bibinfo {author} {\bibfnamefont {R.}~\bibnamefont
  {Yu}}, \bibinfo {author} {\bibfnamefont {W.}~\bibnamefont {Zhang}}, \bibinfo
  {author} {\bibfnamefont {H.-J.}\ \bibnamefont {Zhang}}, \bibinfo {author}
  {\bibfnamefont {S.-C.}\ \bibnamefont {Zhang}}, \bibinfo {author}
  {\bibfnamefont {X.}~\bibnamefont {Dai}}, \ and\ \bibinfo {author}
  {\bibfnamefont {Z.}~\bibnamefont {Fang}},\ }\href@noop {} {\bibfield
  {journal} {\bibinfo  {journal} {Science}\ }\textbf {\bibinfo {volume}
  {329}},\ \bibinfo {pages} {61} (\bibinfo {year} {2010})}\BibitemShut
  {NoStop}%
\bibitem [{\citenamefont {Zhang}\ \emph {et~al.}(2014)\citenamefont {Zhang},
  \citenamefont {Wang}, \citenamefont {Xu}, \citenamefont {Xu},\ and\
  \citenamefont {Zhang}}]{haijunCdOEuO}%
  \BibitemOpen
  \bibfield  {author} {\bibinfo {author} {\bibfnamefont {H.}~\bibnamefont
  {Zhang}}, \bibinfo {author} {\bibfnamefont {J.}~\bibnamefont {Wang}},
  \bibinfo {author} {\bibfnamefont {G.}~\bibnamefont {Xu}}, \bibinfo {author}
  {\bibfnamefont {Y.}~\bibnamefont {Xu}}, \ and\ \bibinfo {author}
  {\bibfnamefont {S.-C.}\ \bibnamefont {Zhang}},\ }\href {\doibase
  10.1103/PhysRevLett.112.096804} {\bibfield  {journal} {\bibinfo  {journal}
  {Phys. Rev. Lett.}\ }\textbf {\bibinfo {volume} {112}},\ \bibinfo {pages}
  {096804} (\bibinfo {year} {2014})}\BibitemShut {NoStop}%
\bibitem [{\citenamefont {Xu}\ \emph {et~al.}(2011)\citenamefont {Xu},
  \citenamefont {Weng}, \citenamefont {Wang}, \citenamefont {Dai},\ and\
  \citenamefont {Fang}}]{xu2011chern}%
  \BibitemOpen
  \bibfield  {author} {\bibinfo {author} {\bibfnamefont {G.}~\bibnamefont
  {Xu}}, \bibinfo {author} {\bibfnamefont {H.}~\bibnamefont {Weng}}, \bibinfo
  {author} {\bibfnamefont {Z.}~\bibnamefont {Wang}}, \bibinfo {author}
  {\bibfnamefont {X.}~\bibnamefont {Dai}}, \ and\ \bibinfo {author}
  {\bibfnamefont {Z.}~\bibnamefont {Fang}},\ }\href@noop {} {\bibfield
  {journal} {\bibinfo  {journal} {Physical review letters}\ }\textbf {\bibinfo
  {volume} {107}},\ \bibinfo {pages} {186806} (\bibinfo {year}
  {2011})}\BibitemShut {NoStop}%
\bibitem [{\citenamefont {Chang}\ \emph {et~al.}(2013)\citenamefont {Chang},
  \citenamefont {Zhang}, \citenamefont {Feng}, \citenamefont {Shen},
  \citenamefont {Zhang}, \citenamefont {Guo}, \citenamefont {Li}, \citenamefont
  {Ou}, \citenamefont {Wei}, \citenamefont {Wang}, \citenamefont {Ji},
  \citenamefont {Feng}, \citenamefont {Ji}, \citenamefont {Chen}, \citenamefont
  {Jia}, \citenamefont {Dai}, \citenamefont {Fang}, \citenamefont {Zhang},
  \citenamefont {He}, \citenamefont {Wang}, \citenamefont {Lu}, \citenamefont
  {Ma},\ and\ \citenamefont {Xue}}]{chang2013experimental}%
  \BibitemOpen
  \bibfield  {author} {\bibinfo {author} {\bibfnamefont {C.-Z.}\ \bibnamefont
  {Chang}}, \bibinfo {author} {\bibfnamefont {J.}~\bibnamefont {Zhang}},
  \bibinfo {author} {\bibfnamefont {X.}~\bibnamefont {Feng}}, \bibinfo {author}
  {\bibfnamefont {J.}~\bibnamefont {Shen}}, \bibinfo {author} {\bibfnamefont
  {Z.}~\bibnamefont {Zhang}}, \bibinfo {author} {\bibfnamefont
  {M.}~\bibnamefont {Guo}}, \bibinfo {author} {\bibfnamefont {K.}~\bibnamefont
  {Li}}, \bibinfo {author} {\bibfnamefont {Y.}~\bibnamefont {Ou}}, \bibinfo
  {author} {\bibfnamefont {P.}~\bibnamefont {Wei}}, \bibinfo {author}
  {\bibfnamefont {L.-L.}\ \bibnamefont {Wang}}, \bibinfo {author}
  {\bibfnamefont {Z.-Q.}\ \bibnamefont {Ji}}, \bibinfo {author} {\bibfnamefont
  {Y.}~\bibnamefont {Feng}}, \bibinfo {author} {\bibfnamefont {S.}~\bibnamefont
  {Ji}}, \bibinfo {author} {\bibfnamefont {X.}~\bibnamefont {Chen}}, \bibinfo
  {author} {\bibfnamefont {J.}~\bibnamefont {Jia}}, \bibinfo {author}
  {\bibfnamefont {X.}~\bibnamefont {Dai}}, \bibinfo {author} {\bibfnamefont
  {Z.}~\bibnamefont {Fang}}, \bibinfo {author} {\bibfnamefont {S.-C.}\
  \bibnamefont {Zhang}}, \bibinfo {author} {\bibfnamefont {K.}~\bibnamefont
  {He}}, \bibinfo {author} {\bibfnamefont {Y.}~\bibnamefont {Wang}}, \bibinfo
  {author} {\bibfnamefont {L.}~\bibnamefont {Lu}}, \bibinfo {author}
  {\bibfnamefont {X.-C.}\ \bibnamefont {Ma}}, \ and\ \bibinfo {author}
  {\bibfnamefont {Q.-K.}\ \bibnamefont {Xue}},\ }\href@noop {} {\bibfield
  {journal} {\bibinfo  {journal} {Science}\ }\textbf {\bibinfo {volume}
  {340}},\ \bibinfo {pages} {167} (\bibinfo {year} {2013})}\BibitemShut
  {NoStop}%
\bibitem [{\citenamefont {Halperin}(1987)}]{halperin1987possible}%
  \BibitemOpen
  \bibfield  {author} {\bibinfo {author} {\bibfnamefont {B.~I.}\ \bibnamefont
  {Halperin}},\ }\href@noop {} {\bibfield  {journal} {\bibinfo  {journal}
  {Japanese Journal of Applied Physics}\ }\textbf {\bibinfo {volume} {26}},\
  \bibinfo {pages} {1913} (\bibinfo {year} {1987})}\BibitemShut {NoStop}%
\bibitem [{\citenamefont {Hyart}\ \emph {et~al.}(2017)\citenamefont {Hyart},
  \citenamefont {Ojaj{\"a}rvi},\ and\ \citenamefont
  {Heikkil{\"a}}}]{hyart2017two}%
  \BibitemOpen
  \bibfield  {author} {\bibinfo {author} {\bibfnamefont {T.}~\bibnamefont
  {Hyart}}, \bibinfo {author} {\bibfnamefont {R.}~\bibnamefont {Ojaj{\"a}rvi}},
  \ and\ \bibinfo {author} {\bibfnamefont {T.}~\bibnamefont {Heikkil{\"a}}},\
  }\href@noop {} {\bibfield  {journal} {\bibinfo  {journal} {arXiv preprint
  arXiv:1709.05265}\ } (\bibinfo {year} {2017})}\BibitemShut {NoStop}%
\bibitem [{\citenamefont {Blaha}\ \emph {et~al.}()\citenamefont {Blaha},
  \citenamefont {Schwarz}, \citenamefont {Madsen}, \citenamefont {Kvasnicka},\
  and\ \citenamefont {Luitz}}]{blaha2002wien2k}%
  \BibitemOpen
  \bibfield  {author} {\bibinfo {author} {\bibfnamefont {P.}~\bibnamefont
  {Blaha}}, \bibinfo {author} {\bibfnamefont {K.}~\bibnamefont {Schwarz}},
  \bibinfo {author} {\bibfnamefont {G.}~\bibnamefont {Madsen}}, \bibinfo
  {author} {\bibfnamefont {D.}~\bibnamefont {Kvasnicka}}, \ and\ \bibinfo
  {author} {\bibfnamefont {J.}~\bibnamefont {Luitz}},\ }\href@noop {} {\bibinfo
   {journal} {WIEN2k, An Augmented Plane Wave Plus Local Orbitals Program for
  Calculating Crystal Properties (TU Vienna, Vienna, 2001)}\ }\BibitemShut
  {NoStop}%
\bibitem [{\citenamefont {Kresse}\ and\ \citenamefont
  {Furthm{\"u}ller}(1996{\natexlab{a}})}]{kresse1996efficiency}%
  \BibitemOpen
\bibfield  {journal} {  }\bibfield  {author} {\bibinfo {author} {\bibfnamefont
  {G.}~\bibnamefont {Kresse}}\ and\ \bibinfo {author} {\bibfnamefont
  {J.}~\bibnamefont {Furthm{\"u}ller}},\ }\href@noop {} {\bibfield  {journal}
  {\bibinfo  {journal} {Computational materials science}\ }\textbf {\bibinfo
  {volume} {6}},\ \bibinfo {pages} {15} (\bibinfo {year}
  {1996}{\natexlab{a}})}\BibitemShut {NoStop}%
\bibitem [{\citenamefont {Kresse}\ and\ \citenamefont
  {Furthm{\"u}ller}(1996{\natexlab{b}})}]{kresse1996efficient}%
  \BibitemOpen
  \bibfield  {author} {\bibinfo {author} {\bibfnamefont {G.}~\bibnamefont
  {Kresse}}\ and\ \bibinfo {author} {\bibfnamefont {J.}~\bibnamefont
  {Furthm{\"u}ller}},\ }\href@noop {} {\bibfield  {journal} {\bibinfo
  {journal} {Physical review B}\ }\textbf {\bibinfo {volume} {54}},\ \bibinfo
  {pages} {11169} (\bibinfo {year} {1996}{\natexlab{b}})}\BibitemShut {NoStop}%
\bibitem [{\citenamefont {Perdew}\ and\ \citenamefont {Wang}(1992)}]{LDA}%
  \BibitemOpen
  \bibfield  {author} {\bibinfo {author} {\bibfnamefont {J.~P.}\ \bibnamefont
  {Perdew}}\ and\ \bibinfo {author} {\bibfnamefont {Y.}~\bibnamefont {Wang}},\
  }\href {\doibase 10.1103/PhysRevB.45.13244} {\bibfield  {journal} {\bibinfo
  {journal} {Phys. Rev. B}\ }\textbf {\bibinfo {volume} {45}},\ \bibinfo
  {pages} {13244} (\bibinfo {year} {1992})}\BibitemShut {NoStop}%
\bibitem [{\citenamefont {Liechtenstein}\ \emph {et~al.}(1995)\citenamefont
  {Liechtenstein}, \citenamefont {Anisimov},\ and\ \citenamefont
  {Zaanen}}]{liechtenstein1995density}%
  \BibitemOpen
  \bibfield  {author} {\bibinfo {author} {\bibfnamefont {A.}~\bibnamefont
  {Liechtenstein}}, \bibinfo {author} {\bibfnamefont {V.}~\bibnamefont
  {Anisimov}}, \ and\ \bibinfo {author} {\bibfnamefont {J.}~\bibnamefont
  {Zaanen}},\ }\href@noop {} {\bibfield  {journal} {\bibinfo  {journal}
  {Physical Review B}\ }\textbf {\bibinfo {volume} {52}},\ \bibinfo {pages}
  {R5467} (\bibinfo {year} {1995})}\BibitemShut {NoStop}%
\bibitem [{\citenamefont {Marzari}\ \emph {et~al.}(2012)\citenamefont
  {Marzari}, \citenamefont {Mostofi}, \citenamefont {Yates}, \citenamefont
  {Souza},\ and\ \citenamefont {Vanderbilt}}]{marzari2012maximally}%
  \BibitemOpen
  \bibfield  {author} {\bibinfo {author} {\bibfnamefont {N.}~\bibnamefont
  {Marzari}}, \bibinfo {author} {\bibfnamefont {A.~A.}\ \bibnamefont
  {Mostofi}}, \bibinfo {author} {\bibfnamefont {J.~R.}\ \bibnamefont {Yates}},
  \bibinfo {author} {\bibfnamefont {I.}~\bibnamefont {Souza}}, \ and\ \bibinfo
  {author} {\bibfnamefont {D.}~\bibnamefont {Vanderbilt}},\ }\href@noop {}
  {\bibfield  {journal} {\bibinfo  {journal} {Reviews of Modern Physics}\
  }\textbf {\bibinfo {volume} {84}},\ \bibinfo {pages} {1419} (\bibinfo {year}
  {2012})}\BibitemShut {NoStop}%
\bibitem [{\citenamefont {Sancho}\ \emph {et~al.}(1984)\citenamefont {Sancho},
  \citenamefont {Sancho},\ and\ \citenamefont {Rubio}}]{sancho1984quick}%
  \BibitemOpen
  \bibfield  {author} {\bibinfo {author} {\bibfnamefont {M.~L.}\ \bibnamefont
  {Sancho}}, \bibinfo {author} {\bibfnamefont {J.~L.}\ \bibnamefont {Sancho}},
  \ and\ \bibinfo {author} {\bibfnamefont {J.}~\bibnamefont {Rubio}},\
  }\href@noop {} {\bibfield  {journal} {\bibinfo  {journal} {Journal of Physics
  F: Metal Physics}\ }\textbf {\bibinfo {volume} {14}},\ \bibinfo {pages}
  {1205} (\bibinfo {year} {1984})}\BibitemShut {NoStop}%
\bibitem [{\citenamefont {Sancho}\ \emph {et~al.}(1985)\citenamefont {Sancho},
  \citenamefont {Sancho}, \citenamefont {Sancho},\ and\ \citenamefont
  {Rubio}}]{sancho1985highly}%
  \BibitemOpen
  \bibfield  {author} {\bibinfo {author} {\bibfnamefont {M.~L.}\ \bibnamefont
  {Sancho}}, \bibinfo {author} {\bibfnamefont {J.~L.}\ \bibnamefont {Sancho}},
  \bibinfo {author} {\bibfnamefont {J.~L.}\ \bibnamefont {Sancho}}, \ and\
  \bibinfo {author} {\bibfnamefont {J.}~\bibnamefont {Rubio}},\ }\href@noop {}
  {\bibfield  {journal} {\bibinfo  {journal} {Journal of Physics F: Metal
  Physics}\ }\textbf {\bibinfo {volume} {15}},\ \bibinfo {pages} {851}
  (\bibinfo {year} {1985})}\BibitemShut {NoStop}%
\bibitem [{\citenamefont {Wu}\ \emph {et~al.}(2018)\citenamefont {Wu},
  \citenamefont {Zhang}, \citenamefont {Song}, \citenamefont {Troyer},\ and\
  \citenamefont {Soluyanov}}]{wu2017wanniertools}%
  \BibitemOpen
  \bibfield  {author} {\bibinfo {author} {\bibfnamefont {Q.}~\bibnamefont
  {Wu}}, \bibinfo {author} {\bibfnamefont {S.}~\bibnamefont {Zhang}}, \bibinfo
  {author} {\bibfnamefont {H.-F.}\ \bibnamefont {Song}}, \bibinfo {author}
  {\bibfnamefont {M.}~\bibnamefont {Troyer}}, \ and\ \bibinfo {author}
  {\bibfnamefont {A.~A.}\ \bibnamefont {Soluyanov}},\ }\href {\doibase
  https://doi.org/10.1016/j.cpc.2017.09.033} {\bibfield  {journal} {\bibinfo
  {journal} {Computer Physics Communications}\ }\textbf {\bibinfo {volume}
  {224}},\ \bibinfo {pages} {405 } (\bibinfo {year} {2018})}\BibitemShut
  {NoStop}%
\bibitem [{\citenamefont {Kane}\ and\ \citenamefont
  {Mele}(2005)}]{kane2005quantum}%
  \BibitemOpen
  \bibfield  {author} {\bibinfo {author} {\bibfnamefont {C.~L.}\ \bibnamefont
  {Kane}}\ and\ \bibinfo {author} {\bibfnamefont {E.~J.}\ \bibnamefont
  {Mele}},\ }\href@noop {} {\bibfield  {journal} {\bibinfo  {journal} {Physical
  review letters}\ }\textbf {\bibinfo {volume} {95}},\ \bibinfo {pages}
  {226801} (\bibinfo {year} {2005})}\BibitemShut {NoStop}%
\bibitem [{\citenamefont {Bernevig}\ \emph {et~al.}(2006)\citenamefont
  {Bernevig}, \citenamefont {Hughes},\ and\ \citenamefont
  {Zhang}}]{bernevig2006quantum}%
  \BibitemOpen
  \bibfield  {author} {\bibinfo {author} {\bibfnamefont {B.~A.}\ \bibnamefont
  {Bernevig}}, \bibinfo {author} {\bibfnamefont {T.~L.}\ \bibnamefont
  {Hughes}}, \ and\ \bibinfo {author} {\bibfnamefont {S.-C.}\ \bibnamefont
  {Zhang}},\ }\href@noop {} {\bibfield  {journal} {\bibinfo  {journal}
  {Science}\ }\textbf {\bibinfo {volume} {314}},\ \bibinfo {pages} {1757}
  (\bibinfo {year} {2006})}\BibitemShut {NoStop}%
\bibitem [{\citenamefont {K{\"o}nig}\ \emph {et~al.}(2007)\citenamefont
  {K{\"o}nig}, \citenamefont {Wiedmann}, \citenamefont {Br{\"u}ne},
  \citenamefont {Roth}, \citenamefont {Buhmann}, \citenamefont {Molenkamp},
  \citenamefont {Qi},\ and\ \citenamefont {Zhang}}]{konig2007quantum}%
  \BibitemOpen
  \bibfield  {author} {\bibinfo {author} {\bibfnamefont {M.}~\bibnamefont
  {K{\"o}nig}}, \bibinfo {author} {\bibfnamefont {S.}~\bibnamefont {Wiedmann}},
  \bibinfo {author} {\bibfnamefont {C.}~\bibnamefont {Br{\"u}ne}}, \bibinfo
  {author} {\bibfnamefont {A.}~\bibnamefont {Roth}}, \bibinfo {author}
  {\bibfnamefont {H.}~\bibnamefont {Buhmann}}, \bibinfo {author} {\bibfnamefont
  {L.~W.}\ \bibnamefont {Molenkamp}}, \bibinfo {author} {\bibfnamefont {X.-L.}\
  \bibnamefont {Qi}}, \ and\ \bibinfo {author} {\bibfnamefont {S.-C.}\
  \bibnamefont {Zhang}},\ }\href@noop {} {\bibfield  {journal} {\bibinfo
  {journal} {Science}\ }\textbf {\bibinfo {volume} {318}},\ \bibinfo {pages}
  {766} (\bibinfo {year} {2007})}\BibitemShut {NoStop}%
\bibitem [{\citenamefont {Fu}\ \emph {et~al.}(2007)\citenamefont {Fu},
  \citenamefont {Kane},\ and\ \citenamefont {Mele}}]{fu2007topological}%
  \BibitemOpen
  \bibfield  {author} {\bibinfo {author} {\bibfnamefont {L.}~\bibnamefont
  {Fu}}, \bibinfo {author} {\bibfnamefont {C.~L.}\ \bibnamefont {Kane}}, \ and\
  \bibinfo {author} {\bibfnamefont {E.~J.}\ \bibnamefont {Mele}},\ }\href@noop
  {} {\bibfield  {journal} {\bibinfo  {journal} {Physical review letters}\
  }\textbf {\bibinfo {volume} {98}},\ \bibinfo {pages} {106803} (\bibinfo
  {year} {2007})}\BibitemShut {NoStop}%
\bibitem [{\citenamefont {Nielsen}\ and\ \citenamefont
  {Ninomiya}(1981{\natexlab{a}})}]{nielsen1981absence1}%
  \BibitemOpen
  \bibfield  {author} {\bibinfo {author} {\bibfnamefont {H.~B.}\ \bibnamefont
  {Nielsen}}\ and\ \bibinfo {author} {\bibfnamefont {M.}~\bibnamefont
  {Ninomiya}},\ }\href@noop {} {\bibfield  {journal} {\bibinfo  {journal}
  {Nuclear Physics B}\ }\textbf {\bibinfo {volume} {185}},\ \bibinfo {pages}
  {20} (\bibinfo {year} {1981}{\natexlab{a}})}\BibitemShut {NoStop}%
\bibitem [{\citenamefont {Nielsen}\ and\ \citenamefont
  {Ninomiya}(1981{\natexlab{b}})}]{nielsen1981absence2}%
  \BibitemOpen
  \bibfield  {author} {\bibinfo {author} {\bibfnamefont {H.~B.}\ \bibnamefont
  {Nielsen}}\ and\ \bibinfo {author} {\bibfnamefont {M.}~\bibnamefont
  {Ninomiya}},\ }\href@noop {} {\bibfield  {journal} {\bibinfo  {journal}
  {Nuclear Physics B}\ }\textbf {\bibinfo {volume} {193}},\ \bibinfo {pages}
  {173} (\bibinfo {year} {1981}{\natexlab{b}})}\BibitemShut {NoStop}%
\bibitem [{\citenamefont {Fang}\ and\ \citenamefont {Fu}(2015)}]{fang2015new}%
  \BibitemOpen
  \bibfield  {author} {\bibinfo {author} {\bibfnamefont {C.}~\bibnamefont
  {Fang}}\ and\ \bibinfo {author} {\bibfnamefont {L.}~\bibnamefont {Fu}},\
  }\href@noop {} {\bibfield  {journal} {\bibinfo  {journal} {Physical Review
  B}\ }\textbf {\bibinfo {volume} {91}},\ \bibinfo {pages} {161105} (\bibinfo
  {year} {2015})}\BibitemShut {NoStop}%
\bibitem [{\citenamefont {Shiozaki}\ \emph {et~al.}(2015)\citenamefont
  {Shiozaki}, \citenamefont {Sato},\ and\ \citenamefont
  {Gomi}}]{shiozaki2015z}%
  \BibitemOpen
  \bibfield  {author} {\bibinfo {author} {\bibfnamefont {K.}~\bibnamefont
  {Shiozaki}}, \bibinfo {author} {\bibfnamefont {M.}~\bibnamefont {Sato}}, \
  and\ \bibinfo {author} {\bibfnamefont {K.}~\bibnamefont {Gomi}},\ }\href@noop
  {} {\bibfield  {journal} {\bibinfo  {journal} {Physical Review B}\ }\textbf
  {\bibinfo {volume} {91}},\ \bibinfo {pages} {155120} (\bibinfo {year}
  {2015})}\BibitemShut {NoStop}%
\bibitem [{\citenamefont {Fang}\ and\ \citenamefont
  {Fu}(2017)}]{fang2017rotation}%
  \BibitemOpen
  \bibfield  {author} {\bibinfo {author} {\bibfnamefont {C.}~\bibnamefont
  {Fang}}\ and\ \bibinfo {author} {\bibfnamefont {L.}~\bibnamefont {Fu}},\
  }\href@noop {} {\bibfield  {journal} {\bibinfo  {journal} {arXiv preprint
  arXiv:1709.01929}\ } (\bibinfo {year} {2017})}\BibitemShut {NoStop}%
\bibitem [{\citenamefont {Araujo}\ and\ \citenamefont
  {Corbett}(1981)}]{araujo1981lanthanum}%
  \BibitemOpen
  \bibfield  {author} {\bibinfo {author} {\bibfnamefont {R.~E.}\ \bibnamefont
  {Araujo}}\ and\ \bibinfo {author} {\bibfnamefont {J.~D.}\ \bibnamefont
  {Corbett}},\ }\href@noop {} {\bibfield  {journal} {\bibinfo  {journal}
  {Inorganic Chemistry}\ }\textbf {\bibinfo {volume} {20}},\ \bibinfo {pages}
  {3082} (\bibinfo {year} {1981})}\BibitemShut {NoStop}%
\bibitem [{\citenamefont {Mattausch}\ \emph {et~al.}(1980)\citenamefont
  {Mattausch}, \citenamefont {Simon}, \citenamefont {Holzer},\ and\
  \citenamefont {Eger}}]{mattausch1980monohalogenide}%
  \BibitemOpen
  \bibfield  {author} {\bibinfo {author} {\bibfnamefont {H.}~\bibnamefont
  {Mattausch}}, \bibinfo {author} {\bibfnamefont {A.}~\bibnamefont {Simon}},
  \bibinfo {author} {\bibfnamefont {N.}~\bibnamefont {Holzer}}, \ and\ \bibinfo
  {author} {\bibfnamefont {R.}~\bibnamefont {Eger}},\ }\href@noop {} {\bibfield
   {journal} {\bibinfo  {journal} {Zeitschrift f{\"u}r anorganische und
  allgemeine Chemie}\ }\textbf {\bibinfo {volume} {466}},\ \bibinfo {pages} {7}
  (\bibinfo {year} {1980})}\BibitemShut {NoStop}%
\bibitem [{\citenamefont {Weng}\ \emph {et~al.}(2014)\citenamefont {Weng},
  \citenamefont {Dai},\ and\ \citenamefont {Fang}}]{weng2014transition}%
  \BibitemOpen
  \bibfield  {author} {\bibinfo {author} {\bibfnamefont {H.}~\bibnamefont
  {Weng}}, \bibinfo {author} {\bibfnamefont {X.}~\bibnamefont {Dai}}, \ and\
  \bibinfo {author} {\bibfnamefont {Z.}~\bibnamefont {Fang}},\ }\href@noop {}
  {\bibfield  {journal} {\bibinfo  {journal} {Physical review X}\ }\textbf
  {\bibinfo {volume} {4}},\ \bibinfo {pages} {011002} (\bibinfo {year}
  {2014})}\BibitemShut {NoStop}%
\bibitem [{\citenamefont {Dion}(2004)}]{dion2004m}%
  \BibitemOpen
  \bibfield  {author} {\bibinfo {author} {\bibfnamefont {M.}~\bibnamefont
  {Dion}},\ }\href@noop {} {\bibfield  {journal} {\bibinfo  {journal} {Phys.
  Rev. Lett.}\ }\textbf {\bibinfo {volume} {92}},\ \bibinfo {pages} {246401}
  (\bibinfo {year} {2004})}\BibitemShut {NoStop}%
\bibitem [{\citenamefont {Gong}\ \emph {et~al.}(2017)\citenamefont {Gong},
  \citenamefont {Li}, \citenamefont {Li}, \citenamefont {Ji}, \citenamefont
  {Stern}, \citenamefont {Xia}, \citenamefont {Cao}, \citenamefont {Bao},
  \citenamefont {Wang}, \citenamefont {Wang}, \citenamefont {Qiu},
  \citenamefont {Cava}, \citenamefont {Louie}, \citenamefont {Xia},\ and\
  \citenamefont {Zhang}}]{gong2017discovery}%
  \BibitemOpen
  \bibfield  {author} {\bibinfo {author} {\bibfnamefont {C.}~\bibnamefont
  {Gong}}, \bibinfo {author} {\bibfnamefont {L.}~\bibnamefont {Li}}, \bibinfo
  {author} {\bibfnamefont {Z.}~\bibnamefont {Li}}, \bibinfo {author}
  {\bibfnamefont {H.}~\bibnamefont {Ji}}, \bibinfo {author} {\bibfnamefont
  {A.}~\bibnamefont {Stern}}, \bibinfo {author} {\bibfnamefont
  {Y.}~\bibnamefont {Xia}}, \bibinfo {author} {\bibfnamefont {T.}~\bibnamefont
  {Cao}}, \bibinfo {author} {\bibfnamefont {W.}~\bibnamefont {Bao}}, \bibinfo
  {author} {\bibfnamefont {C.}~\bibnamefont {Wang}}, \bibinfo {author}
  {\bibfnamefont {Y.}~\bibnamefont {Wang}}, \bibinfo {author} {\bibfnamefont
  {Z.~Q.}\ \bibnamefont {Qiu}}, \bibinfo {author} {\bibfnamefont {R.~J.}\
  \bibnamefont {Cava}}, \bibinfo {author} {\bibfnamefont {S.~G.}\ \bibnamefont
  {Louie}}, \bibinfo {author} {\bibfnamefont {J.}~\bibnamefont {Xia}}, \ and\
  \bibinfo {author} {\bibfnamefont {X.}~\bibnamefont {Zhang}},\ }\href@noop {}
  {\bibfield  {journal} {\bibinfo  {journal} {Nature}\ } (\bibinfo {year}
  {2017})}\BibitemShut {NoStop}%
\bibitem [{\citenamefont {Slater}\ and\ \citenamefont {Koster}(1954)}]{sktb}%
  \BibitemOpen
  \bibfield  {author} {\bibinfo {author} {\bibfnamefont {J.~C.}\ \bibnamefont
  {Slater}}\ and\ \bibinfo {author} {\bibfnamefont {G.~F.}\ \bibnamefont
  {Koster}},\ }\href {\doibase 10.1103/PhysRev.94.1498} {\bibfield  {journal}
  {\bibinfo  {journal} {Phys. Rev.}\ }\textbf {\bibinfo {volume} {94}},\
  \bibinfo {pages} {1498} (\bibinfo {year} {1954})}\BibitemShut {NoStop}%
\bibitem [{\citenamefont {Sarma}\ \emph {et~al.}(2011)\citenamefont {Sarma},
  \citenamefont {Adam}, \citenamefont {Hwang},\ and\ \citenamefont
  {Rossi}}]{sarma2011electronic}%
  \BibitemOpen
  \bibfield  {author} {\bibinfo {author} {\bibfnamefont {S.~D.}\ \bibnamefont
  {Sarma}}, \bibinfo {author} {\bibfnamefont {S.}~\bibnamefont {Adam}},
  \bibinfo {author} {\bibfnamefont {E.}~\bibnamefont {Hwang}}, \ and\ \bibinfo
  {author} {\bibfnamefont {E.}~\bibnamefont {Rossi}},\ }\href@noop {}
  {\bibfield  {journal} {\bibinfo  {journal} {Reviews of Modern Physics}\
  }\textbf {\bibinfo {volume} {83}},\ \bibinfo {pages} {407} (\bibinfo {year}
  {2011})}\BibitemShut {NoStop}%
\end{thebibliography}

%

\end{document}